\documentclass{he_symp}
\usepackage{psfig}
\usepackage{graphics}

\def \bold{\bf}
\def\etal{{\it et al.\/}}
\def\eg{{\it e.g.\/}}

\def \Epar{{\mathbf E}_{\parallel}}
\def\cf{{\it cf.\/}}
\def\ie{{\it i.e.\/}}
\def\ltw{\>\hbox{\lower.25em\hbox{$\buildrel <\over\sim$}}\>}
\def\gtw{\>\hbox{\lower.25em\hbox{$\buildrel >\over\sim$}}\>}

\def \Om{\Omega}
\def \g{\gamma}
\def \del{\nabla}
\def \be{\begin{equation}}
\def \ee{\end{equation}}

\begin{document}

\title{The Radio-Loud Plasma in Pulsars}

\author{J. A. Eilek\inst{1}, P. N. Arendt, Jr.\inst{1}, T. H. Hankins\inst{1}
 \and J. C. Weatherall\inst{1}}

\institute{New Mexico Tech, Socorro NM 87801, USA} 

\maketitle

\begin{abstract}

The pulsar magnetosphere contains a strongly magnetized, relativistic
plasma.  We need to understand the physics of that plasma if we want
to connect the data to the models.  Our group in Socorro is mixing
 theory and observations in order to study the radio-loud pulsar
plasma.  In this paper we report on several aspects of our current work. 

\end{abstract}

\section{Introduction}

We know pulsars produce radio emission at high brightness temperatures.
It is very likely that this arises in a plasma. We also know that
 high-$T_{\rm B}$ 
emission is created by dynamical, non-equilibrium plasma processes
elsewhere in space or astrophysical plasmas.   Therefore, we can
safely guess that the process by which pulsars shine involves a plasma.
This is, however, not saying much.  We know very little about the origin or
dynamics of the radio-loud plasma. This is unfortunate, because
understanding the pulsar plasma is critical to understanding the star. 
The radiation from that plasma is the only data we
can get from these interesting stars.   If we want to connect the observations
to the physics (including the wealth of new, high-quality data becoming
available, such as presented at this meeting), we need to study the
plasma.

This paper is organized around two questions.  What do we know
about the radio-loud plasma in the pulsar?   Can insight and 
methods from other areas of plasma astrophysics be applied here?
More specifically, can we understand the dynamical state of the plasma
(which connects back to the electrodynamics of the system), and 
how it produces the observed
radiation (which is a consequence of its dynamical state)?

We begin with an overview (\S 2-5) of standard pulsar
theory and what that theory requires, or suggests, 
about the plasma.  One common assumption is that the radio emission
comes from a lepton pair plasma.  In \S 6-7 we report on work carried out
by two of us (Arendt \& Eilek) which follows  the pair cascade
numerically and determines the details of the lepton plasma 
produced by the cascade.  After that, in \S 8-9 we suggest that
time variability might be the best observational diagnostic of the
plasma, and we report on ongoing work by two of us 
(Weatherall, Eilek) to model the plasma dynamics on short
timescales.  We close with a report (\S 10) on ultra-high
time resolution observations of the Crab pulsar carried out by one 
of us (Hankins) 
which we believe begin to address some basic theoretical issues
regarding plasma dynamics and the emission mechanism.

\section{The standard model and its Limitations}

To set the stage, we review the standard model,
with an eye to its consequences for plasma formation and dynamics.
This model has been carefully developed by many people over the years,
and has had good success.  But it also has its limitations;  now
that we have a common theoretical ground, and now that we have
some excellent new data (including single-pulse work, polarimetry,
multifrequency studies and millisecond pulsar studies), it is
 time to review the model with a critical eye.  
The areas where it fails will be the areas which can be fruitfully
revisited.  In this paper we limit ourselves to the radio emission
region, and leave high-altitude, high-energy issues to others.

\subsection{A corotating magnetosphere}

A strong magnetic field is tied to a rapidly rotating neutron star.  It
contains and structures a plasma-filled magnetosphere.  Most of the field
lines which leave the star's surface reverse direction within the light
cylinder and return to the star (as illustrated in Figure 1).
 It is  natural to consider a steady
atmosphere in this region, \ie, one which rotates with the star.
This can be maintained if an electric field exists, 
${\mathbf E}_{\rm co}$, which satisfies
\be
{\mathbf E}_{\rm co} + { 1 \over c} 
\left( {\mathbf \Omega} \times {\mathbf r} \right) 
\times {\mathbf B} = 0
\ee
Such a field produces an  $\mathbf{E}_{\rm co} \times
\mathbf{B}$ drift equal to the corotation speed.  This field must be
supported by a nonzero charge density, 
\be
\rho_{\rm GJ} = { 1 \over 4 \pi} \nabla \cdot {\mathbf E}_{\rm co}
\simeq - {\mathbf{\Om} \cdot \mathbf{B} \over 2 \pi c}
\ee
 This is called the ``corotation'' or ``GJ'' 
density.  Note that the plasma must be non-neutral in order to
corotate;  most of the magnetosphere is usually assumed to contain a
single sign of charge. 

\vspace{-0.1in}
\begin{figure}[htb]
\centerline{\psfig{file=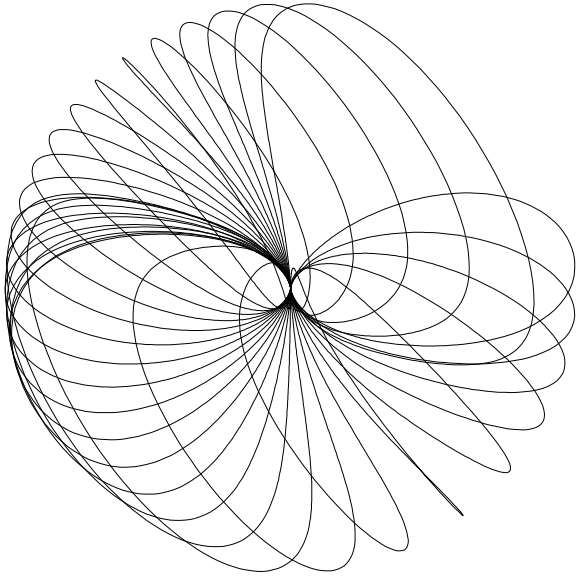,width=9.5cm,clip=} }
\vspace{-0.2in}
\caption{The magnetosphere of a pulsar, illustrated by the
last open field lines in a rotating vacuum magnetic dipole,
at an $45^{\circ}$ inclination angle to the rotation axis, 
viewed looking down the rotation axis. Relativistic effects 
distort the field lines close to the light cylinder.  
Most of the magnetosphere, within the closed field lines, 
is assumed to corotate with the star.  The open field
line region (or polar flux tube, PFT)
allows a current-carrying outflow from the star,  is
thought to contain a dense electron-positron plasma, and to be the
origin of the radio emission. Despite the high-altitude field
distortions due to relativity, the PFT is nearly circular in cross
section at radio emission altitudes.   From Arendt (2002).
}
\end{figure}

\subsection{The acceleration region}

Conditions are different in the polar flux tube (PFT).  This is
the narrow region defined
by field lines which do not close within the light cylinder. 
In this  region, relativistic outflow from star's surface is thought
to occur, driven by deviations from the 
corotation density.  These deviations can support an $\bold E$ field
 with a non-zero 
component parallel to the $\bold B$ field; $\Epar$ can  accelerate
particles to high energies.

The standard model now has a branch point, with two specific but
different variants proposed (\eg, Ruderman \& Sutherland 1975, Arons
\& Scharleman 1979).   One variant assumes that charges cannot
freely leave the star's surface, and thus a ``vacuum gap''
exists just above the surface.  Strong $\Epar$
in this gap accelerates particles to relativistic energies, perhaps
in a non-steady fashion (sparking, breakdown).  
Another variant assumes that charges can freely leave the star's
surface.  If the charge density deviates from corotation, 
$\Epar$ created by these charges will enable their
acceleration, also to relativistic energies.

This low-altitude acceleration creates a highly relativistic ``primary
beam'', which gains significant fraction of the maximum potential drop in a
short distance.  Most authors assume the acceleration region is limited
by pair formation, which is believed to shield $\Epar$.  Alternatively,
the acceleration may be limited by radiative losses before the
energies required for pair formation are reached (Sturmer 1995; Jessner
\etal\/ 2001). 

\subsection{The Pair Cascade}

A very important step is taken here.  Radiation from the primary
beam charges is assumed to seed a $\gamma B$ pair cascade.
Either curvature radiation (if the charges reach $\gamma \sim 10^7$)
or magnetic-resonant Compton scattering 
(if they reach $\gamma \sim 10^4$ and the star is
a thermal X-ray source) produces  energetic photons.   These
photons, in the star's strong $B$ field,
can produce an electron-positron pair.  Further photon production
by these daughter leptons leads to further pair production,
thus initiating a cascade of particle creation.  The end point of
this cascade is often assumed to be total conversion of the primary
beam energy to much slower (but still relativistic) leptons, with
very high multiplicities (${\cal M}$, defined as the number of
daughter leptons per primary beam particle).  

Two additional important assumptions follow;  neither of them has
been clearly established to be true.  One is that the pair
plasma terminates the acceleration region, by shielding the accelerating
$E$ field.   The second is that
the  pair plasma is the means by which the free energy
available in the primary beam is converted to coherent radio emission.
We discuss both in more detail later. 

\section{Annoying Issues and Inconsistencies}

Some problems remain in this picture. 

\subsection{The Density is a Problem}

We do not know the details of the radio emission mechanism, but
the local plasma frequency, $\nu_{\rm p}$, is likely to be important.
Plasma turbulent models predict emission at close to $\nu_{\rm p}$. 
Signal propagation is suppressed below $\nu_{\rm p}$ (except at angles
very close to $\bold B$).  However, the numbers do not work.
(Others have also noticed this problem:  \cf\/ Kunzl \etal\/ 1998, or
Melrose 2000). 
  
Consider a plasma of number density $n$, moving at bulk Lorentz
factor $\gamma$.  It supports plasma waves with frequency
$\nu_{\rm p}^2 = \gamma n e^2 / \pi m$ (all quantities measured in
the star's frame).  The standard 
model predicts the number density is at least that required
by corotation, $n_{\rm GJ} = \rho_{\rm GJ}/e$;  pair creation amplifies
this by the multiplicity ${\cal M}$.  If we use (2) and scale
to the field at the stellar surface,  this predicts a plasma frequency
\be
\nu_{\rm p} = \left( { \gamma e B {\cal M} \over 
4 \pi m c P } \right)^{1/2} \simeq 1.2 \left[{ \gamma B_{*,12} {\cal M}
\over P ( r /  r_* )^3 } \right]^{1/2} {\rm GHz}
\ee
Now, $\gamma {\cal M} \gg 1$ in most models, and $r 
\sim 3-30 r_*$ is the typical range of emission altitudes. 
Thus, the standard  model predicts 
$\nu_{\rm p} > 1$ GHz, and we should see little radiation at 
lower frequencies.  On the other hand,
pulsars are strong radio sources down to frequencies $\sim 10$ MHz.
We consider this a problem. 

To resolve this problem,
either the radiation must emerge well below the plasma frequency,
or the plasma density in at least part of the
radio-loud region is much lower than
commonly thought.  We think the latter case is more likely.  It
must follow that steady corotation is not obeyed; the plasma will
be dynamically variable.

\subsection{The Plasma is not Steady}

We note two further hints of an unsteady plasma. 

No pulsar has steady radio emission.  We see modulation on all time
scales where we've looked, from sub-$\mu$s flickering to mode changes
and nulls on multi-period timescales.  Coherent
radio emission is a very nonlinear-gain process, so we 
expect  small fluctuations
in the plasma to be strongly amplified.  Nonetheless, this observed
variability tells us that the plasma 
flow cannot be as steady as in most of the models.  
As with the density problem, 
this suggests to us that the plasma  does not satisfy cororation. 
Density deviations lead to a net force in the corotating frame, and
these forces should lead to nonsteady flows. 

Furthermore, the assumption of a steady, corotating magnetosphere
with no flows  is only
self-consistent for an aligned rotator.  We note  two interesting
features of a star whose magnetic axis is not aligned with its
rotation axis.  
(1) The $\bold E \times \bold B$ drift velocity has a component
along $\bold B$.  Unless the plasma is charge-neutral (recall $\rho_{\rm GJ}$
carries a sign), this will drive field-aligned currents even in 
corotating regions.  
(2) The ``GJ current'', $j_{\rm GJ} \propto  \rho_{\rm GJ}$, has a non-zero
divergeance.  Non-steady flows are therefore 
 required by charge conservation, unless the magnetosphere is fully
static (which is highly unlikely). 
Both of these effects may be small, but they
also hint at  non-steady plasma flows in the magnetosphere.
 
\subsection{Where is the acceleration region?}

The standard model predicts a very low-altitude acceleration and pair
cascade region.  On the other hand, the radio-loud region is thought
to be at least several stellar radii above the surface. 
The emission altitude
is determined from the data through a quasi-universal set of assumptions:
that the radio emission arises only in the PFT, that the magnetic
field is dipolar,  and that the mean profile
width is due to the star rotating past one's sightline.  Details of
the deprojection vary between authors (\eg,  Rankin 1993, Kijak \& Gil 1998,
or Eilek \& Hankins 2002),
but there is general agreement that the emission altitude lies between
a few and a few tens of stellar radii.  

We find this unsatisfying.  Once the pair plasma is formed it
should radiate almost immediately.  The plasma turbulence
thought to lead to radio emission should develop quickly. 
Thus, if the acceleration
and pair formation region is indeed close to the star, the 
radiation must remain confined in the PFT plasma until it escapes
at a few tens of stellar radii.  Plasma trapping of the radiation,
when the wave frequency is below the local plasma
frequency, has of course been assumed to explain this.  However, we now
know that the density distribution of the PFT is not as simple as
was once thought. We also note that radiation wavevectors at small 
angles to $\bold B$ can propagate below the local plasma frequency.  
The entire situation would be  simpler if the radio emission region 
were also the acceleration  and  pair formation region (\eg, Shibata
1991).

\subsection{Do all pulsars have a pair cascade? }

The pair cascade is an important part of the standard model.  Its
onset is determined by the acceleration rate in the ``gap'' region.  The
dense plasma it produces is assumed to shield the accelerating $\mathbf{E}$
field, and thus end the gap.  But this cartoon is not self-consistent.
When the standard model is made quantitative, 
one finds that the pair cascade is unlikely to occur in pulsars with
magnetic field $\ltw 10^{11}$G (Arendt \& Eilek 2002; 
details below).  This means
that neither old, isolated pulsars, nor msec pulsars, should show
pair formation,  if they obey the standard model. 
 
How can this be resolved?  Either (i)  not all pulsars have a dense
 pair atmosphere (and thus are ``pairless'' radiators), or (ii) our 
standard model is seriously lacking; or perhaps (iii) both of
the above are true. 

\subsection{Is Magnetic Field Dipolar?}

The assumption that the magnetic field is dipolar lies at the heart of
the standard model.  Detailed development of that model usually
assumes a purely dipolar field.  There are various hints,
however, that this is not the full picture.  Some authors have
suggested that the low-altitude field is more complicated -- a
stronger field with more complex structure near the surface may
help the problem of pair formation in weak-field pulsars. 
Such ideas may well be supported by the data.  There has been
some very good work on polarization 
in recent years (\eg,  van Hoensbroech \& Xilouris 1997,
or  Hankins \& Rankin 2002)
We now know (\eg, Eilek \& Hankins 2000) that the linear polarization in
an interesting subset of bright pulsars is {\it not} well explained
by the usual rotating-vector model.  Thus, the  field is very likely
to deviate from dipolar, {\it even in the radio emission region}.

\section{Larger Unsolved Questions}

The standard picture contains (at least) two major, unsolved questions.
These are not new questions, and work continues on each -- with no
consensus as to the answer.

\subsection{The Pulsar Circuit}

An outflow of relativistic charged particles, along open field lines, is
a fundamental part of the pulsar model.  This constitutes a current.
The nature of the pulsar circuit is still far from understood:  where
does the current return to the star?  Is the circuit open (to the ISM) 
or closed?  How is the current driven (that is,
how does the rotational potential of the star couple to the current
 flow?)  The literature contains two approaches to this problem.  
Models of the low-altitude acceleration region (possibly with pair
formation) describe in detail the dynamics of individual charges
in that region.  This region is assumed also to drive the current. If
the outflowing plasma is relativistic, fills the PFT, and is 
 at the corotation density, then the
magnitude of the current follows directly:  $I = \rho_{\rm GJ} c A_{PFT}$.
These models rarely address the rest of the circuit. 

An alternative approach is to consider the global circuit.  The dynamo
must still be the star's rotation, which creates a potential drop 
in any plasma which is not corotating with the star.\footnote{This may be
 the magnetosphere, which  some authors assume rotates at a slightly 
different rate from the star itself, or could be the ambient ISM, which
is magnetically connected to the star {\it via} the open field lines.}
 This approach allows the 
possibility of localized resistive regions (``loads''; possibly the
inner or outer gaps) where
energy is dumped into plasma, then radiation (\eg, Kunzl \etal\/ 2002).
In principle, the combination of the driving 
voltage and the resistance of the circuit will determine the magnitude
of the current. We see no  {\it a priori} reason for this to be
the GJ current (Shibata, 1991, has made this point nicely.).  

These models have not yet been developed at the level of
the low-altitude acceleration region models, but we find them
attractive.  One might hope that  observational diagnostics could
be developed (of such quantities as current density or circuit
length).  We return to this issue in \S8-9.

\subsection{The Radiation Mechanism}

How does the pulsar shine? We do not have a simple model of how
the plasma produces the
very high brightness temperatures characteristic of the radio emission.
Most authors believe  either the primary beam or the pair plasma
will be susceptible to plasma instabilities (such as beam bunching
or the two-stream instability).  Some of these instabilities
will grow to nonlinearity and couple to electromagnetic radiation which
can escape the system. 

Within this general picture much variety is possible. Models abound,
and some have been developed in great detail, but with no consensus as
to which is the most relevant to real pulsars, and with hardly any
useful observational discriminants proposed. (For instance, \cf\/
Melrose 1992, or Lyubarski, this meeting).  We  might group
the models currently being discussed into three categories.  One 
starts with {\it coherent charge bunches}. Early 
emission models assumed that such bunches produced curvature emission
in the radio range.  
An interesting recent variant is Compton scattering of
plasma emission by such bunches (\eg, Kunzl \etal\/ 2002). 
A second class of models invoke strong {\it plasma turbulence},
in which soliton growth and possible collapse convert 
plasma wave energy into escaping radiation ({\eg ,
Asseo \etal\/ 1990, Weatherall 1997). 
A third type of model uses {\it maser-like} 
mechanisms, which  convert internal energy  of the plasma to radio 
emission (\eg, Luo \& Melrose 1995, Kazbegi \etal\. 1991, Weatherall 2001) 

All of these models differ in their details, and often in their
deductions about the significant properties of the emission (such as
its spectrum, spatial location, or relation to local plasma conditions). 
One might also hope that observational tests of the models
could be developed.
We also return to this question in \S8-9, below.

\section{The State of the Radio-Loud Plasma }  

To return to our original question:  what do we know, or assume,
about the plasma in the radio-loud region?  

\paragraph{Composition.}  The plasma is probably
dominated by electron-positron pairs in 
the radio-loud region (although this is not established for slow and
msec pulsars).  It is also   not  charge neutral (in order
to maintain corotation).  
  
\paragraph{Number density.}  The plasma is  assumed either to be at
the corotation density, $n_{\rm GJ}= \rho_{\rm GJ}/e$, or to be increased over
this by the pair multiplicity ${\cal M}$.  Both of these assumptions
are common in the literature;  as we point out above, both assumptions
are inconsistent with observed low-frequency radio emission.  We
suspect (in agreement with Melrose 2000) that
the plasma is highly inhomogeneous, with radio emission coming
from the lower-density regions, and the corotation density possibly being
met in some average fashion. 

\paragraph{Plasma ``beta''.}  The plasma is highly magnetized, with 
  $B \sim 10^{11} - 10^{13}$G at the surface of
single pulsars. It is therefore strongly field dominated;
the inertia and energy density of the plasma are tiny compared to 
the magnetic field. 

\paragraph{Flows.}  The plasma flow is highly relativistic and
one-dimensional:  $\g \sim 100 - 10^7$
(depending on author).  The flow is
nearly along the magnetic field, with slow cross-field drifts in
regions where transverse $\bold E$ fields exist.  The gyromotion of
individual charges is quantized, with most particles in 
the lowest Landau levels. 

\paragraph{Dynamical state.}  The plasma is very likely to be
inhomogeneous and unsteady.  We know collective effects are 
needed to produce coherent radio emission.  Thus the plasma is
likely to be strongly turbulent on microscales.  We also infer a
nonsteady state from the nonsteady radio emission.  Finally, if
the plasma contains low-density regions, we expect the unshielded
electric field to keep the plasma dynamically active.

\paragraph{Plasma distribution function.}
This important part of any plasma physics study has remained unconstrained
until recently.  Previous authors have either assumed analytically
convenient forms, or tried to quantify semi-heuristic arguments.
New work, on which we report below,  shows that the pair
plasma is well described by a relativistic Maxwellian, 
at $k_{\rm B} T \sim m c^2$, {\it in the comoving frame}.

\section{The Cascade and the Plasma it Produces}

We now turn to the pair cascade:  when and how does it occur, 
and what is the nature of the plasma it produces?  The cascade 
begins with a seed photon,
of energy $\epsilon = h \nu / m c^2$,
 which pair creates {\it via} $\gamma + B \to e^+ + e^- + B$.  The
newly created leptons 
will emit synchrotron radiation if they are created in upper Landau
levels.  If this is the case the synchrotron photons will themselves
pair create, thus extending the cascade to several generations. 

This is an inherently nonlinear process,  not well suited to
analytic methods (although progress can be made with heroic effort,
\eg, Hibschman \& Arons 2001). Numerical simulation are called for.  
Daughtery and Harding (1982) led the way with simulations which
determined the spectrum of hard photons escaping from a cascade
region.  We revisited the subject in order to learn more about 
the pair plasma.
Our goal was to simulate a cascade produced
according to the standard models.  That is, we did not attempt
a self-consistent treatment of the plasma electrodynamics within
the acceleration region.  Rather, we assumed the primary beam
particles had energies typical of what is found in the literature,
and they produce seed energetic photons
either by curvature radiation (CR) or inverse Compton scattering (ICS).
We summarize our results here;  more details can be found in 
Arendt (2002), and  Arendt \& Eilek (2002).  

\subsection{Details of the calculation}

Our calculations were based on monoenergetic photon
``kernels''.  For each of these, we started with photons of a
fixed energy, moving at a small range of angles (such as
given by relativistic beaming) relative to the local $\bold B$.  
We followed each in time and space (assuming linear propagation
relative to a non-rotating frame, ignoring general relativistic
effects) until it had a unit opacity to pair creation.  At that
point the differential cross section and Monte Carlo methods were
 used to create daughter  leptons.  We followed each lepton
in time and space (assuming guiding-center motion along $\bold B$, 
ignoring acceleration by any local $\Epar$)
until it had unit opacity to synchrotron
radiation.  At that point Monte Carlo methods were again used to 
create new photons.  This process continued, following all 
quanta in time and space, until all the action
had stopped---that is until no further pair or photon creation
occured.  The end of such a calculation gave us one ``kernel''.

We convolved the kernels with the seed photon distributions produced
by CR or ICS to determine the plasma and photon distributions
predicted by cascades in a pulsar.  We chose primary beam energies
$\gamma_{\rm p} \sim 10^6 - 10^7$ for cascades seeded by CR,
 and $\gamma_{\rm p} \sim 10^3 - 10^5$ for cascades seeded
by magnetic-resonant ICS.  We simulated full
cascades for magnetic fields $B = 10^{12}$G and $10^{13}$G.
We also explored cascades in $B = 10^{11}$G fields, but found they
were very expensive in computer time, due the large
numbers of soft photons created for each lepton pair produced.  Thus we
only included the two larger field strengths in our final results. 

\subsection{Onset of the Cascade}

The cascade can be understood in terms of the
two governing principles, energetics and opacity. 
Recall that an electron at energy $\gamma_{\rm p}$
produces a curvature radiation spectrum which peaks at a
photon energy  $\epsilon_{\rm CR} = \hbar \omega_{\rm CR} / m c^2 
\sim 3 c \hbar \g^3 / 2 
\rho m c$, where $\rho$ is the radius of curvature of the field line.  
Magnetic resonant ICS produces photons
in the range $\epsilon_{\rm B} / 2 \gamma_{\rm p} \le 
\epsilon \le 2 \gamma_{\rm p}
\epsilon_{\rm B}$, where $\epsilon_{\rm B} = \hbar \omega_{\rm B} / m c^2 
= B / B_{\rm c}$.

Consider a seed photon, of energy $\epsilon$, moving at 
an angle $\theta$ relative to $\bold B$.  Two conditions must
be satisfied if it is to pair produce. 
The first condition is energetics.  The 
available photon energy must satisfy $\epsilon > 2 m c^2$.
Relativistic kinematics expands this as
\be
\epsilon_{\perp} = \epsilon \sin \theta > 2.
\ee
(in the star's frame).  We found that most of the pair creation in
our high-field runs was governed by energetics.
Photons travelling nearly parallel to $\bold B$ propagate freely
until field line curvature finally establishes $\epsilon_{\perp} > 2$.
Pair creation then proceeds immediately.  Most of the
newly created leptons are in the ground Landau state, so they cannot
create further photons, and the cascade ends. 

The second condition that must be met for pair creation is opacity.
The photon must have a significant
chance of creating a pair. Using the basic cross section, one can
show that this requires 
\be
\epsilon B \gtw 0.1 B_{\rm c} \sim 4 \times 10^{12} {\rm G}
\ee
again in the star's frame (this is the best case,
that of $\theta = \pi /2 $, a right angle crossing of the magnetic
field line).   We found that our lower field runs, and some initial
creation events in our high field runs, were opacity dominated.
The parent photons have more than enough energy to
produce a pair, but the probability of pair production is low
until they propagate to an angle such that $\epsilon_{\perp} \ltw
0.1 B_{\rm c}/B$. The newly created leptons in this case tend to
be born in excited Landau levels. They then decay
{\it via} the production of energetic synchrotron photons, which extend
the cascade to several further generations.

These two conditions make it clear that the magnetic pair cascade
does not necessarily occur in all pulsars. 
In particular, stars with a weak $B$ field need very energetic 
photons to satisfy the opacity condition;  such high energies
are not produced by the two standard seed-photon radiation models
(\eg, Weatherall \& Eilek 1997).  It is traditional 
to invoke highly curved fields in old  and
millisecond pulsars, assuming
that this will enable pair production.   We caution,
however,  that the acceleration of the primary beam charges,
and the spectrum of their radiation, must be shown to work
quantitatively before the pair cascade can be demonstrated
to occur in all pulsars. 

\subsection{Development and Decay of the Cascade}

We  found, as expected,  that
pair formation is enhanced towards the edge of the PFT, due to
the larger field line curvature there.  We would thus expect
``conal'' emission from the pair plasma;  as others have
pointed out, ``core'' emission does not easily fit
into this picture. 

We continued our simulations until all reactions had stopped --
until there were no remaining photons with energy and opacity to
create further pairs. 
In all cases the cascade ended before
the quanta reached one stellar radius from their inection point.
Thus, the pair cascade is a rapid process, with the time from seed
photon emission to cascade completion being less than
$R_* / c \sim 30 \mu$s.  If the seed photon production is sporadic,
the pair plasma should be highly nonuniform
within the PFT.

\subsection{Efficiency and Multiplicity of the Cascade}

The fraction of the primary beam energy  converted
into the pair plasma, and the abundance of that plasma, are
key parameters in many models of the pulsar magnetosphere
and its interaction with the ambient ISM.  {\em Our numerical
results do not support assumptions commonly made in the
literature}. 

\begin{table}
\caption{Summary of results from cascade simulations}
\begin{center}
\begin{tabular}{r @{\ $\sim$\ } l r @{\ $\sim$\ } l}
\multicolumn{4}{c}{Curvature Radiation:} \\  
\multicolumn{4}{c}{\ } \\
\multicolumn{2}{c}{$\bold B = 10^{12}$G:} & \multicolumn{2}{c}{$\bold B = 10^{13}$G:}\\
$\gamma_{\rm p}        $&$ 5 - 10 \times 10^6$ & $\gamma_{\rm p}         $&$ 2   - 10 \times 10^6$\\
$ {\cal M}       $&$ 300 - 1000$         & $ {\cal M}        $&$ 500 - 2000$\\
$ f_{e^+e^-}     $&$ 0.03 - 0.1$         & $ f_{e^+e^-}      $&$ 0.4 - 0.5$\\
$ \gamma_{\rm CM}$&$ 80 - 480$           & $ \gamma_{\rm CM} $&$ 60  - 600 $\\
$ f_{\nu}        $&$ 0.02 - 1.0$         & $ f_{\nu}         $&$ 0.001 - 0.03$\\
\multicolumn{4}{c}{\ } \\
\multicolumn{4}{c}{Inverse Compton:} \\
\multicolumn{4}{c}{\ } \\
\multicolumn{2}{c}{$\bold B = 10^{12}$G:} & \multicolumn{2}{c}{$\bold B = 10^{13}$G:}\\
$\gamma_{\rm p}         $&$ 2 - 50 \times 10^4 $&$\gamma_{\rm p}        $&$ 2 - 10 \times 10^4$\\
${\cal M}         $&$ 10 - 100           $&${\cal M}        $&$ 2 - 10$  \\
$f_{e^+e^-}       $&$ 0.1 - 0.3          $&$f_{e^+e^-}      $&$ 0.8 - 1.0$\\
$ \gamma_{\rm CM} $&$ 60 - 400           $&$\gamma_{\rm CM} $&$ 100 - 900 $\\
$ f_{\nu}         $&$ 0.1 - 0.8          $& \multicolumn{2}{c}{$f_{\nu} < 0.01$\quad\quad\quad}\\
\end{tabular}
\end{center}

\noindent
$\gamma_{\rm p}$ is the energy of the primary beam charges.
\\
$\cal M$ is the multiplicity:  the number of pairs per primary charge.
\\
$f_{e^+e^-}$ is the energy efficiency:  the fraction of the primary beam
energy that goes into leptons.
\\
$\gamma_{\rm CM}$ is the center of momentum energy of the created
leptons.
\\
$f_{\nu}$ is the photon
energy efficiency:  the fraction of the primary beam
energy that goes into photons.
\\
Full details can be found in Arendt \& Eilek (2002), or Arendt (2002).
\end{table}

One common assumption is that the entire energy of the primary
beam is transferred to the plasma.  We found this is not always
 the case.  Instead, much of the beam energy escapes as photons. 
The fraction of beam energy going into the pairs ranges from
 $f_{e^+e^-} \ltw 1.0$ in higher magnetic fields, to 
 $f_{e^+e^-} \ltw 0.1$ in lower magnetic fields. 
Another common assumption is that the multiplicity (the number
of pairs produced per primary beam particle) is high;  a value
${\cal M} \sim 10^4$ is often found in the literature.  We
found much lower values.  The high-field runs, with Compton
seeds, were strongly limited by energetics and
 produced only a few pairs per primary charge.  Multiplicities
for the lower field and curvature-seed runs ranged from $O(10)$
to $O(1000)$, and increased strongly with primary beam energy. 
Table 1 summarizes the range of $f_{e^+e^-}$ and $\cal M$ found
in our runs;  more details can be found in the original papers.

\subsection{Lepton Distribution from the Cascade}

One of our primary goals in these simulations was to determine
the momentum distribution function (DF) of the pair plasma
created by the cascade.  We found a very simple result.
Despite the wide range in cascade parameters
and pair creation efficiencies encountered, the final DF's have
remarkably similar shapes.  Figure 2 shows some examples of
DF's seeded by both CR and by ICS, {\em as viewed in the star's
frame}.  The DF's are all broadly similar, with a strong peak
at moderate energies and an extended high-energy tail.  There
are slight differences due to the magnetic field.  In the $B =
10^{13}$G runs, there are no slow leptons;  all DF's have a 
low-$p_{\parallel}$ gap below $\sim 20 m c$.  In contrast, all of
the $B = 10^{12}$G runs have leptons down to $p_{\parallel} < .01
m c$. (This difference is due to the importance of
synchrotron losses, and the energy degrading of
multiple lepton generations, in the lower-field runs). 

\begin{figure}[htb]
\centerline{\psfig{file=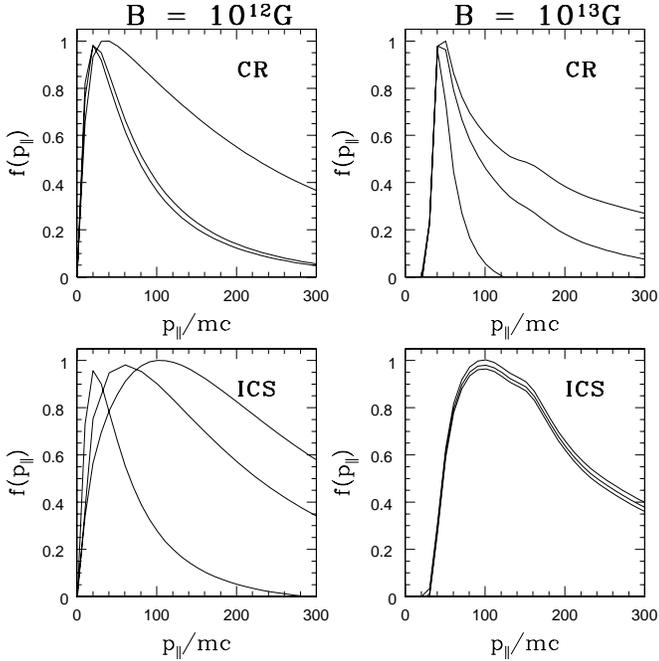,width=9cm,clip=} }
\caption{The one-dimensional momentum distribution function of the pair
plasma created in the cascade, {\it as viewed in the star's frame}.  The
results are shown for two magnetic fields,  for 
cascades seeded by CR with $\gamma_{\rm p} = 1, 2, 5
\times 10^6$, and by ICS with $\gamma_{\rm p} = 2, 5,  10 
\times 10^4$. From Arendt \& Eilek (2002).
\label{image}}
\end{figure}

\begin{figure}[htb]
\centerline{\psfig{file=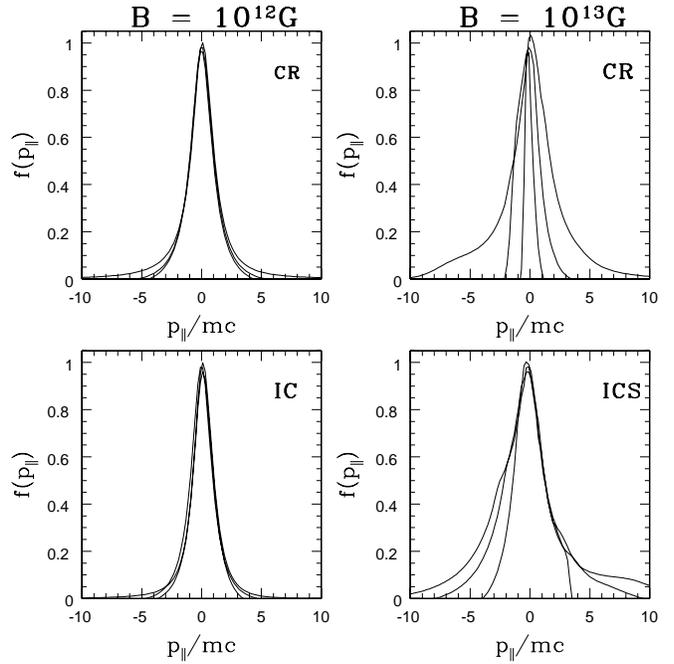,width=9cm,clip=} }
\caption{The one-dimensional momentum DF of the pair
plasma created in the cascade, {\it as viewed in the comoving frame}.  
See Figure 2 for details.   From Arendt \& Eilek (2002).
\label{image}}
\end{figure}

The numerical results simplify  nicely if the DF's
are transformed to the center-of-momentum (CM) frame of the plasma.
(We found that $\gamma_{\rm CM} \sim 10 - 100$ over the set of
runs). 
Figure 3 shows the same DF's as in Figure 2, 
but seen in the CM frame.  In this frame, the DF's
are remarkably close to a {\em thermal} distribution, {\it i.e.\/} 
a relativistic Maxwellian:
\be
f(p_{\parallel}) \propto  \exp \left[- \zeta \gamma
(p_{\parallel}) \right]
\ee
The parameter $\zeta$ can be identified with the inverse temperature
of the plasma.  We carried out numerical fits of this expression
to our CM-frame DF's, and found that $k_{\rm B}
 T/ m c^2  \sim 0.8 - 2$ describes
the full range of our results.  

Thus, the DF of the cascade-produced pair plasma is remarkably
simple.  It is a warm ($k_{\rm B} T \sim m c^2$) relativistic Maxwellian
when viewed in a frame moving with the plasma.  The distorted
shape as seen in the lab is simply due to the Lorentz boost 
when passing to that frame.

\subsection{Photon Distribution from the Cascade}
 
\begin{figure}[htb]
\centerline{\psfig{file=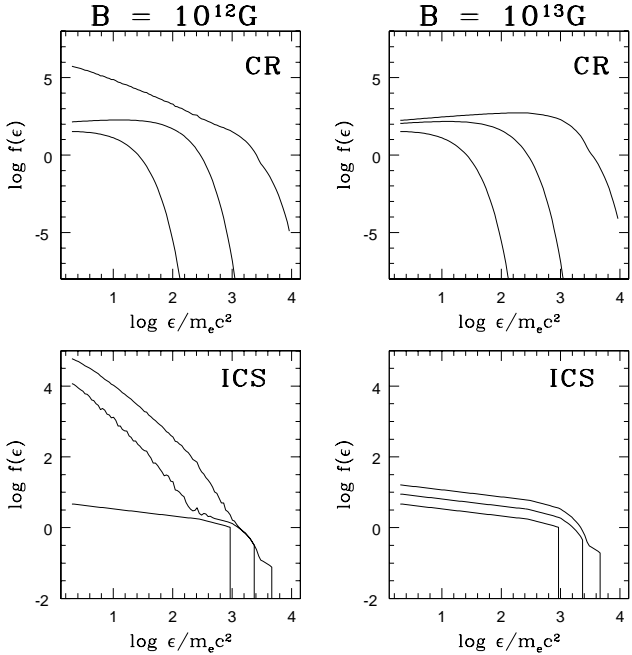,width=9cm,clip=} }
\caption{The photon number counts (not the energy spectrum)
resulting from the pair cascade, in the star's
frame.  See Figure 2 for details.  From Arendt \& Eilek (2002).
\label{image}}
\end{figure}

We also determined the spectrum of the energetic photons which
remained after the end of the cascade.  This is a composite of
secondary photons, produced by synchrotron emission of the
pairs, and seed photons which never pair produced.\footnote{Some photons
never satisfied the opacity and energy criteria.  We distributed
the seed photons over angle, within a small range identified with
$1 / \gamma_{\rm p}$;  those at lower angles had a good chance of
escaping the region without pair producing.}
Figure 4 shows a selection of these spectra, presented as 
number counts (not energy spectra).  The photon spectra tended to
be steeper in lower magnetic fields, due to secondary synchrotron
emission being more important in those cascades.  Conversely,
the spectra are flatter in the high field runs, where synchrotron
photons are rare and the photon spectra are dominated by the
seed photons which escaped pair creation.  All spectra
show a high-energy turnover which comes from the input seed
photon distribution.

The fraction of the primary beam power that remained in the
photons, $f_{\nu}$, was quite low for the high-field runs, especially
those with ICS seed;  nearly all of the beam energy did go into
leptons in those cases. The lower-field runs ended with a
large fraction of the beam energy in photons:  $f_{\nu} \sim 0.1
- 1$ in those runs.  The large numbers of photons produced in
our exploratory runs at $ B = 10^{11}$G suggest $f_{\nu}$ would be
quite large there as well.

\section{Why does the DF matter?}

The reader who is not a plasma theorist may wonder at our emphasis
on the plasma DF.  Our answer is that it is crucial to understanding
the emission and propagation of the radio signal, and that it
may have important consequences for the electrodynamics of the
emission region.

\subsection{Coherent Radio Emission}

As we noted above, the means by which pulsars produce coherent
radio emission are still unknown.  There is a general sense that
some plasma process converts the energy of that beam to radio
emision.  But the details are  critical.  What plasma
process (that is, which instability) is involved? What conditions 
are necessary in the plasma for it to happen?    Where in the 
magnetosphere does it go?   

These questions can only be answered  with a detailed 
knowledge of the plasma DF. Consider the ongoing
discussion about the two-stream instability, and whether or not it
can occur in pulsars. The
instability goes only if the two plasmas moving
relative to each other are are ``cool'',  that is with 
momentum width less than their separation. In the pulsar context, the
two streams are either the electons and positrons of the pair
swarms, or the primary beam moving through the pair swarms.  
The onset of plasma turbulence is governed by the temperature and
relative motion of the different plasma species (\eg, Weatherall 1994; 
Usov, this proceedings.).  In other words,
we need to know the DF. 

But this DF has been unclear;  in fact,
 ``theorist's freedom'' has prevailed.  Some calculations
of plasma wavemodes and instabilities
 have assumed analytically convenient forms (such as boxcars, cold
plasma, or relativistic Maxwellians in the star's frame).   
Other work has attempted
to quantify a plausible, but heuristic, DF introduced by Arons (1981).
We hope that our simulations of the true DF will be of use to future
work in this area.  

\subsection{Magnetospheric Propagation}

The radio emission must travel through the pair plasma, between
its emission point and its escape from the star.  This propagation
can involve ``shunting'' of radiation along $\bold B$, evolution
of the polarization state, and dispersive and scattering effects.
The details of the transport, and observable signatures,
depend critically on the details of the plasma DF.  A familiar 
example is propagation through
the ISM, for which the frequency dependence of scattering and
propagation are well known.  These signatures can easily be derived
from the DF for the ISM, that is a cold, ion-electron plasma, with
plasma frequency well below the frequency of the observed radiation.   
A similar calculation for propagation in a  relativistically moving,
cold pair plasma predicts that dispersion and scattering in
that plasma could be distinguished from ISM propagation at high
frequencies (Eilek, 2002;  also Lyutikov \& Parikh, 2000,
or Petrova \& Lyubarskii, 2000).  We are not aware of any such studies
that include a warm plasma, which we now know to be the case. 

\subsection{Shielding}

Models of the acceleration region usually assume that the pair
plasma can adjust itself to maintain the corotation density, and
thus reduce the local $\bold E$ to zero, as soon as pair formation
goes.  This is a great help mathematically, as it allows for
straightforward analytic solutions in the acceleration region. 
However, 
it may not be supported by more careful consideration of the cascade.  

Such shielding depends on two things:  the multiplicity of the
cascade, and the DF of the pair plasma.  If the plasma is cold,
and if the electrons and positrons are given a large enough
relative streaming velocity (such as a local ${\mathbf E}_{\parallel}$
would create), number conservation leads to a finite charge density
which can satisfy corotation. 
Shibata \etal\/ (1998) go further, and show 
that steady space-charge-limited
flow is possible for a cold plasma  only if the pair
multiplicity is very high.  This is because newly created opposite-sign 
charges change the field structure of the acceleration region, and too few
pairs (too low an ${\cal M}$) cannot support a self-consistent
$\Epar$ field.  Their numerical examples require
much higher ${\cal M}$ values than our simulations predict.  
The issue is further complicated by the creation of warm plasma, as in
our simulations.  Arendt (2002) applied  $\Epar$ fields to
induce relative streaming in the post-cascade
electron and positron distributions. 
He found that quite low fields will, indeed, produce a net
charge density of the opposite sign, which admits
 the possibility of shielding.  However, a somewhat larger 
$E$ field will reverse some of the opposite-sign charges,
and thus enhance the original density, because of charges leaving
the system.  

We are therefore  not confident that the pair cascade will always
adjust to shield the accelerating $\Epar$.  We suspect pair formation
will induce more interesting, non-steady behavior in the plasma. 

\section{ How might we observe the plasma?}

Theories need to be tested by observations.  Is there any way in
which we can observe the pulsar plasma?

\subsection{Indirectly:  radio emission and propagation}

We observe the plasma indirectly {\it via} its radio emission.  In
more conventional astrophysics this allows a simple approach --
incoherent emission is well understood.  The radio power and spectrum 
allow the type of emission (synchrotron, bremsstrahlung, {\it etc.\/}) to be
identified, and the physical conditions (such as density or temperature)
in the emitting region to be determined.

This is not so easy for pulsars.  We do not have a unique 
model for coherent radio emission, let alone a simple one.  The coherent
power is likely to be extremely sensitive to detailed conditions
in the plasma, so that modest changes in, say, local density or DF
will result in large changes in the instantaneous radio power.
Furthermore, the 
detected pulsar signal probably arises from a highly inhomogeneous
region, stratified in density and possibly distributed over altitude.

The situation is further complicated by the fact that the radio signal 
may have traversed an extended region of the 
plasma before escaping into ``space'' and thence to our detectors.
The effects of propagation through the ISM are reasonably
well understood, but
this is not the case for propagation effects within the pulsar
magnetosphere.  

Thus, interpreting  observations of radio power or spectrum
directly in terms of the conditions in the emitting region is
very difficult.  We need alternative methods. 

\subsection{ Quasi-directly:  temporal fluctuations}

Anyone who has looked at pulsar radio data knows that the signal is
intrinsically variable on many time scales. The
variability can be categorized in terms of observables, and arranged
in order of variability timescale. 

\paragraph{Nulls, modes and the mean profile.}
The pulsar signal varies dramatically from pulse to pulse, in an
apparently random fashion, and yet each star has a meaningful and
stable emission profile when averaged over enough rotation periods
(tens to hundreds, depending on the star; \eg, Rathnasree \&
Rankin 1995).  In addition, some stars
show nulls or mode changes which persist for many rotation periods.
Thus a large-scale order clearly exists within the emission region.

\paragraph{Drifting subpulses.}
Some stars show another type of striking long-term order:  drifting
subpulses.  These are easily identifiable features which drift
through the mean profile (\eg, Ramachandran \etal\/ 2002).  In a few
cases they are known to be long-lived, lasting tens of rotation 
periods, and to circulate around the star's magnetic axis (Deshpande
\& Rankin 2001).  These occur in some but not all stars;  we do
not know why. 

\begin{figure*}[htb]
\centerline{\psfig{file=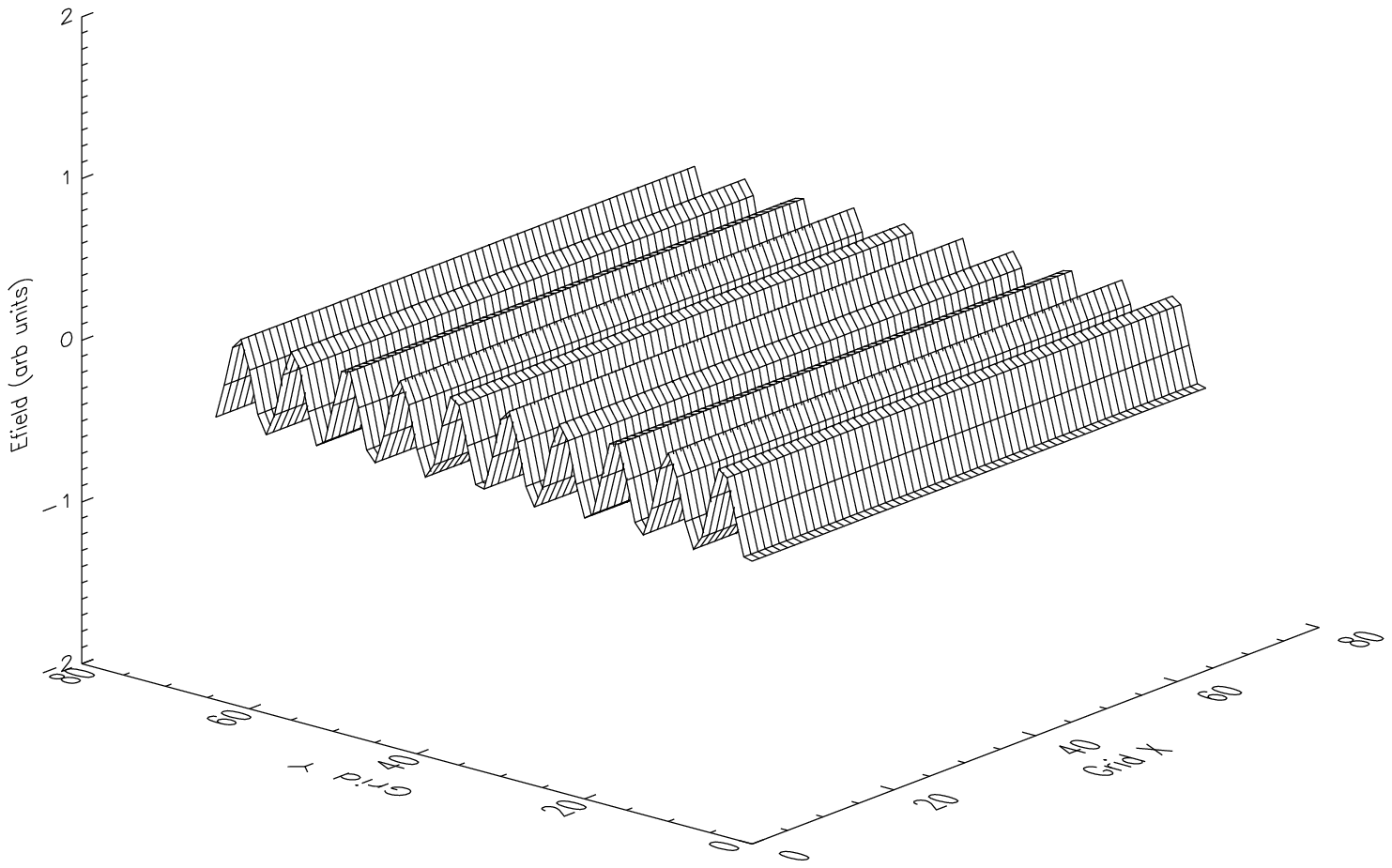,width=6cm,clip=} 
\psfig{file=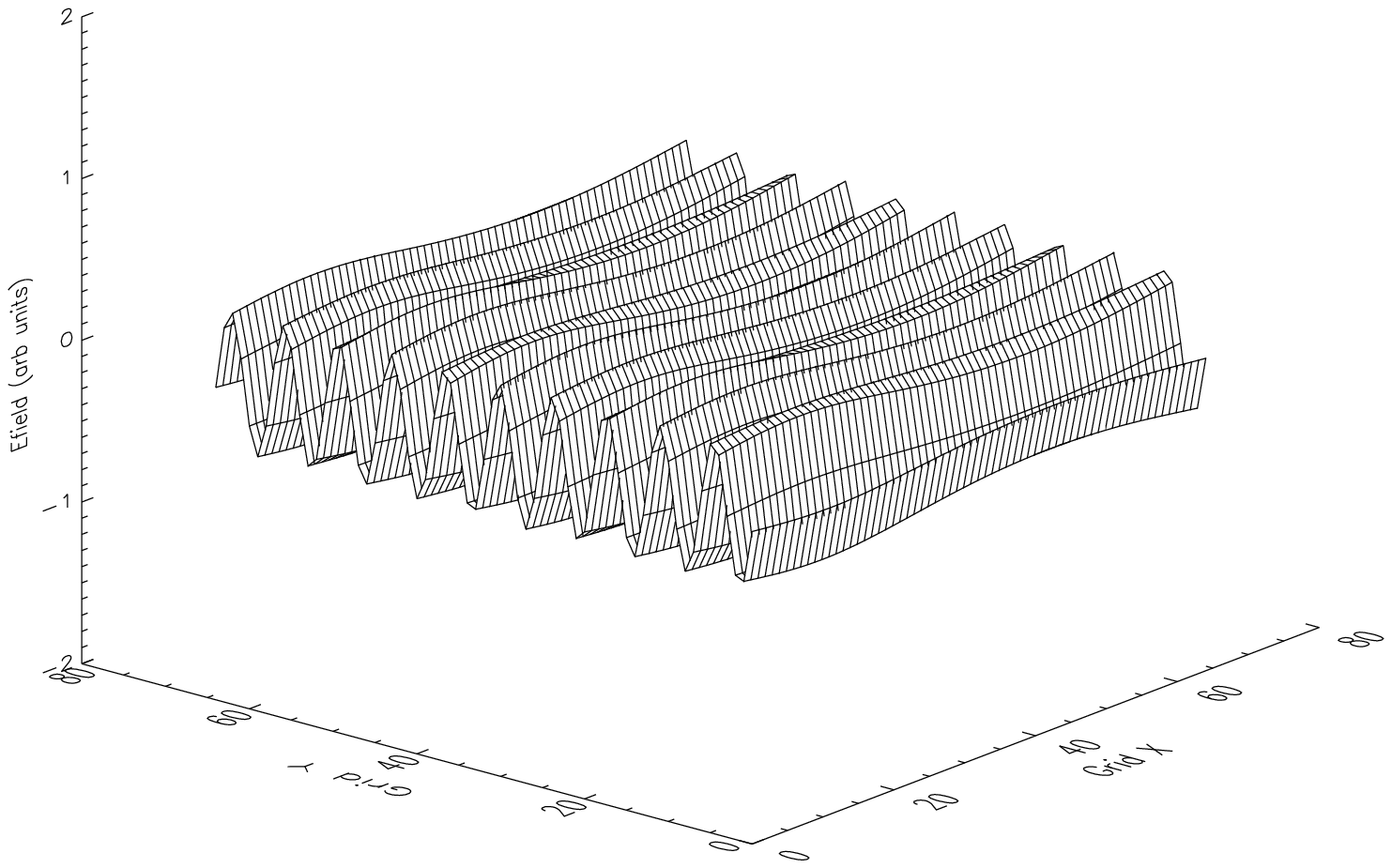,width=6cm,clip=} 
\psfig{file=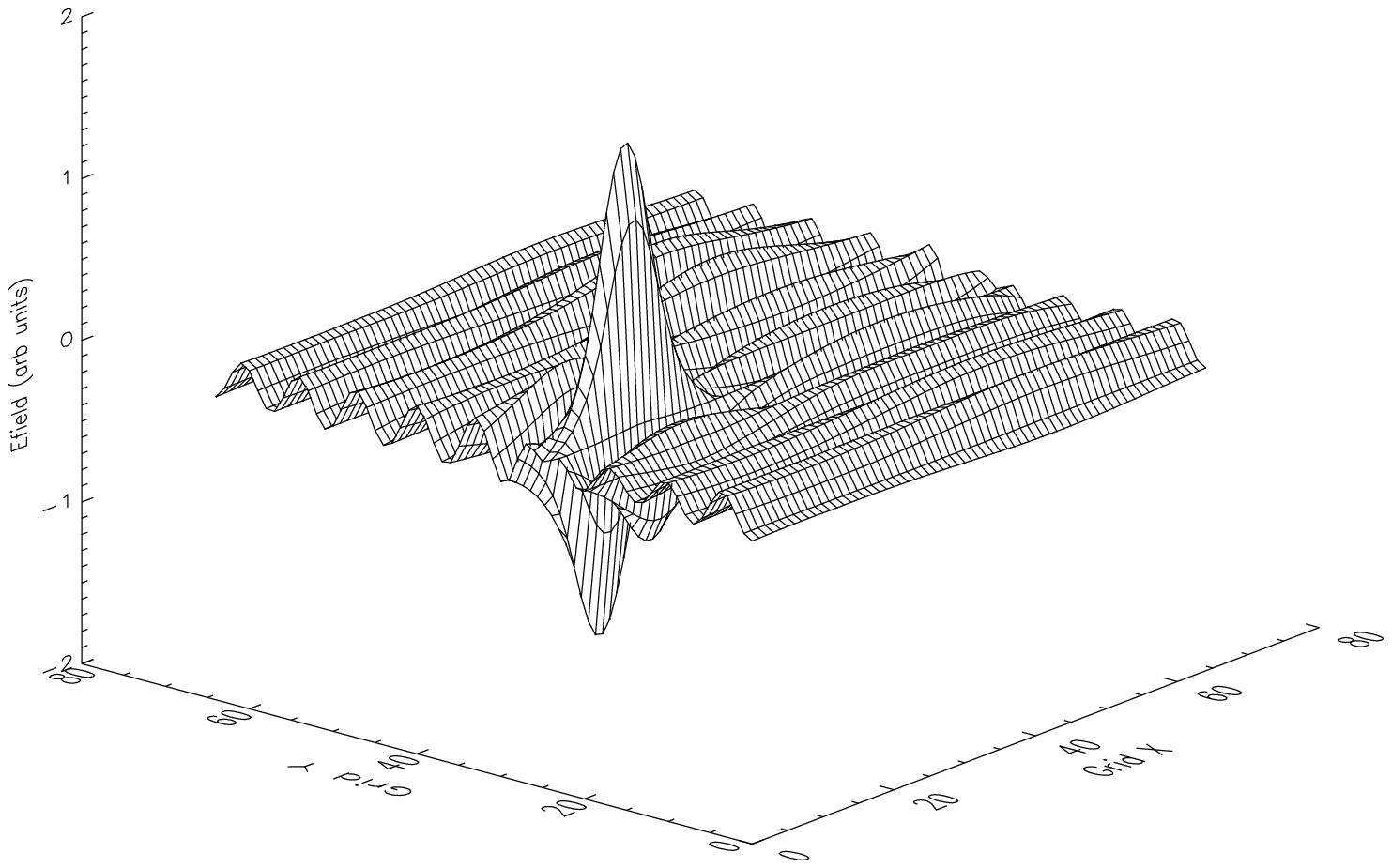,width=6cm,clip=} }
\caption{Numerical simulation of the the electric field in strong
plasma turbulence. The simulation spans a time interval of $200 / \nu_{\rm p}$, 
and illustrates wave growth by beam-instabilty (left), modulational
instability (center), and wavepacket collapse (right).  The extent of
the spatial grid is $ 64 c / \omega_{\rm p}$, which scales to to $60 \gamma$ cm at
an observing frequency of 5 GHz.
The radiation field produced by this collapse is shown in Figure 6.
From Weatherall (1997).   
\label{image}}
\end{figure*}

\paragraph{Microstructure.}
Going to smaller timescales, some stars show 
microstructure, \ie,  significant power on $\delta t \sim 1 - 100
\mu$s (\eg, Lange \etal\/ 1998).   
Once again, microstructure does not exist in all stars (Hankins \& Fowler 
1982, unpublished) and we do not understand why.  
The associated scale size $\ltw
0.3 - 30$ km  $\sim .03 - 3 R_*$. We may be seeing plasma dynamics 
on scales up to the size of the emission region, or we 
may be seeing  a nonsteady pair cascade.

\paragraph{Nanostructure.}
On even smaller time scales, 
nanostructure has been detected in giant pulses from the Crab pulsar.
There is power on $\delta t \sim 10 - 100$ ns. We do not know
whether this occurs in other stars;  only the Crab pulsar has been
studied at this time resolution. The associated spatial 
scale is $\ltw 3 - 30$ m;  we are ``seeing'' quite small structures.

\subsection{Are these temporal or spatial fluctuations?} 

There has been much discussion of whether the short-term variability 
is temporal (due to true time changes in the emitting plasma) or
spatial (as a narrow emission beam moves across the sight line).
We believe it is temporal, and we argue as follows.

Consider the longer timescales.  We can follow individual drifting
subpulses for many rotation periods.  They are surely
due to long-lived plasma structures, moving within the polar cap.
The fact that we can identify these structures as they move around 
the radio emission region   (Deshpande \& Rankin 2001), means that
their emission beam cannot be extremely narrow. 

Compare this to the shortest timescales known. If the Crab nanostructure
were  spatial (as tiny features move past our sightline) the emitting 
plasma would have to  be bunched in  very narrow columns, 
$\ltw 1$ cm in width, 
moving at  bulk Lorentz factors of $\sim 10^7$.  This 
seems unlikely to us, and  contradicts  the evidence from
drifting subpulses.  If microstructure were spatial, the size and
Lorentz factors of the required plasma columns would be less extreme, but
would still disagree with the subpulse evidence.  Based on this, we
believe the short-term variability reflects time variability
in the emitting plasma.  

\section{Dynamical Variability and Timescales -- examples}

We know that other plasmas, in the lab and in space, are dynamic
and highly variable.  We  expect the pulsar plasma
to be so as well. Thus, our observations of variability in the pulsar
signal are indirect observations of the plasma dynamics.  With an
eye to the general models of the pulsar emission region, described above, we
consider  time and space scales on which we  expect
variability. 

\subsection{Micro-scale:  the radiating entities}

On the smallest space and time scales, the plasma variability will 
be that which produces coherent radiation.  As we noted above,
the specific emission process remains a mystery, and several
competing models have been suggested.  
What these models have in common is  sensitivity to the microscale
dynamics of the underlying plasma.  The large-scale structure of
the radio-loud region sets the broad stage, but coherent radio emission
is generated by evolving plasma structures on much
smaller scales.  We suspect that each emission process will have
a characteristic timescale, which can in principle be observed.

\begin{figure}[htb]
\vspace{-0.5in}
\centerline{\psfig{file=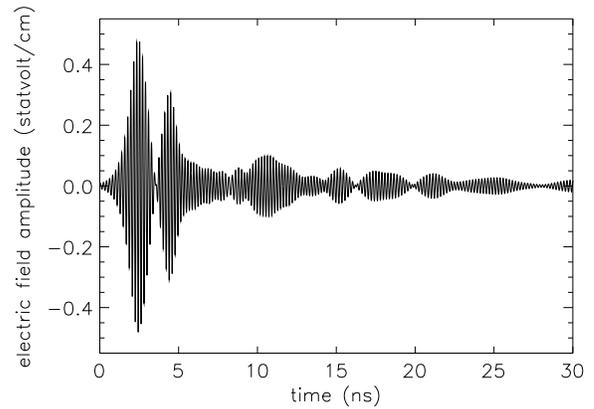,width=9.5cm,clip=} }
\vspace{-2.0in}
\caption{Numerical simulation of a nanopulse:  the amplitude of the
radiation burst emitted by the strong-turbulence region shown in Figure
5.   The characteristic time for radiation emitted at $\nu_o$ 
is $\nu_o \delta t \sim O( 10)$;  this figure has been scaled to an observing
frequency of 5 GHz.  From Weatherall (1998). 
\label{image}}
\end{figure}

\begin{figure*}[htb]
\centerline{
\psfig{file=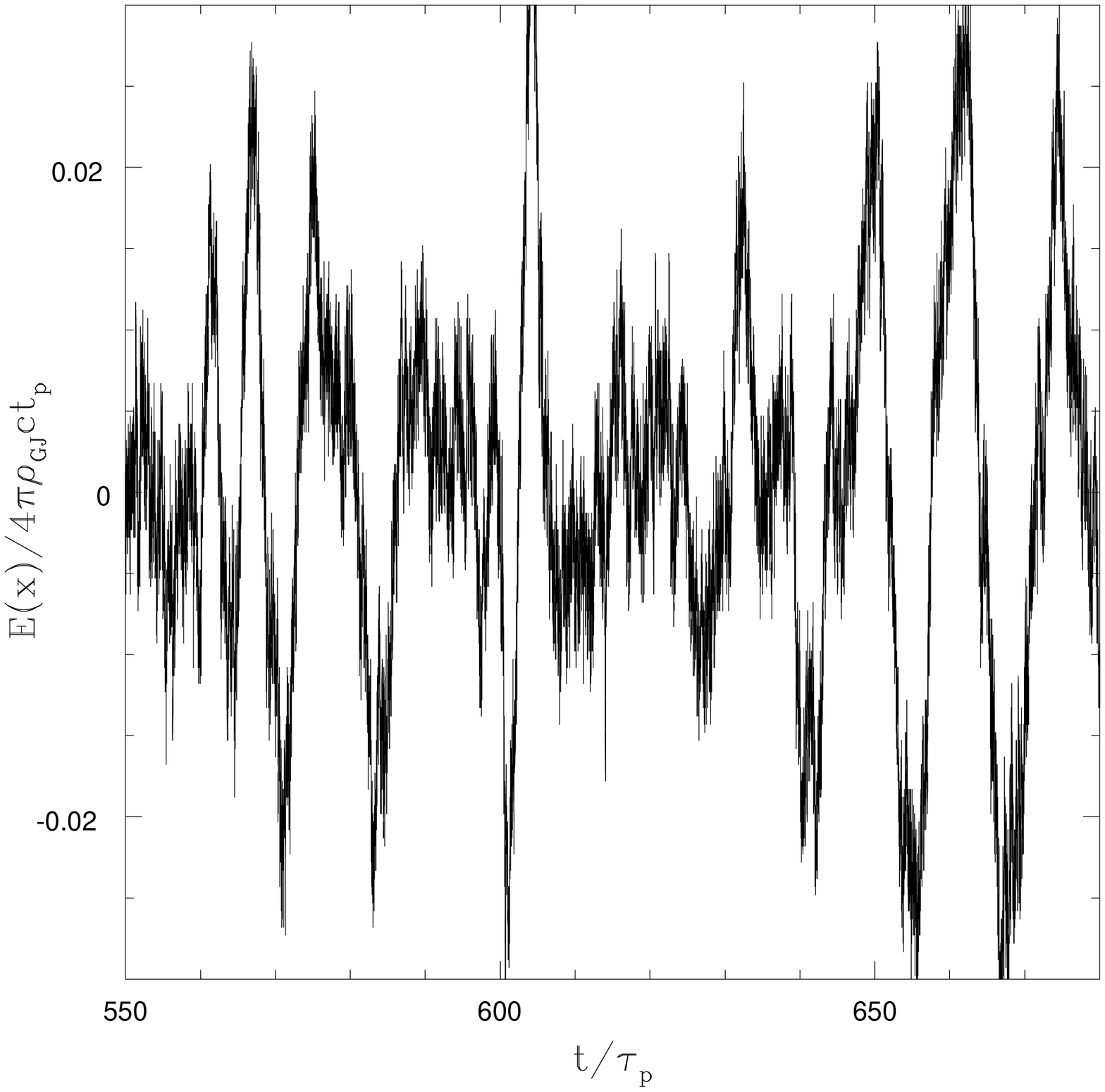,width=6cm,clip=} 
\psfig{file=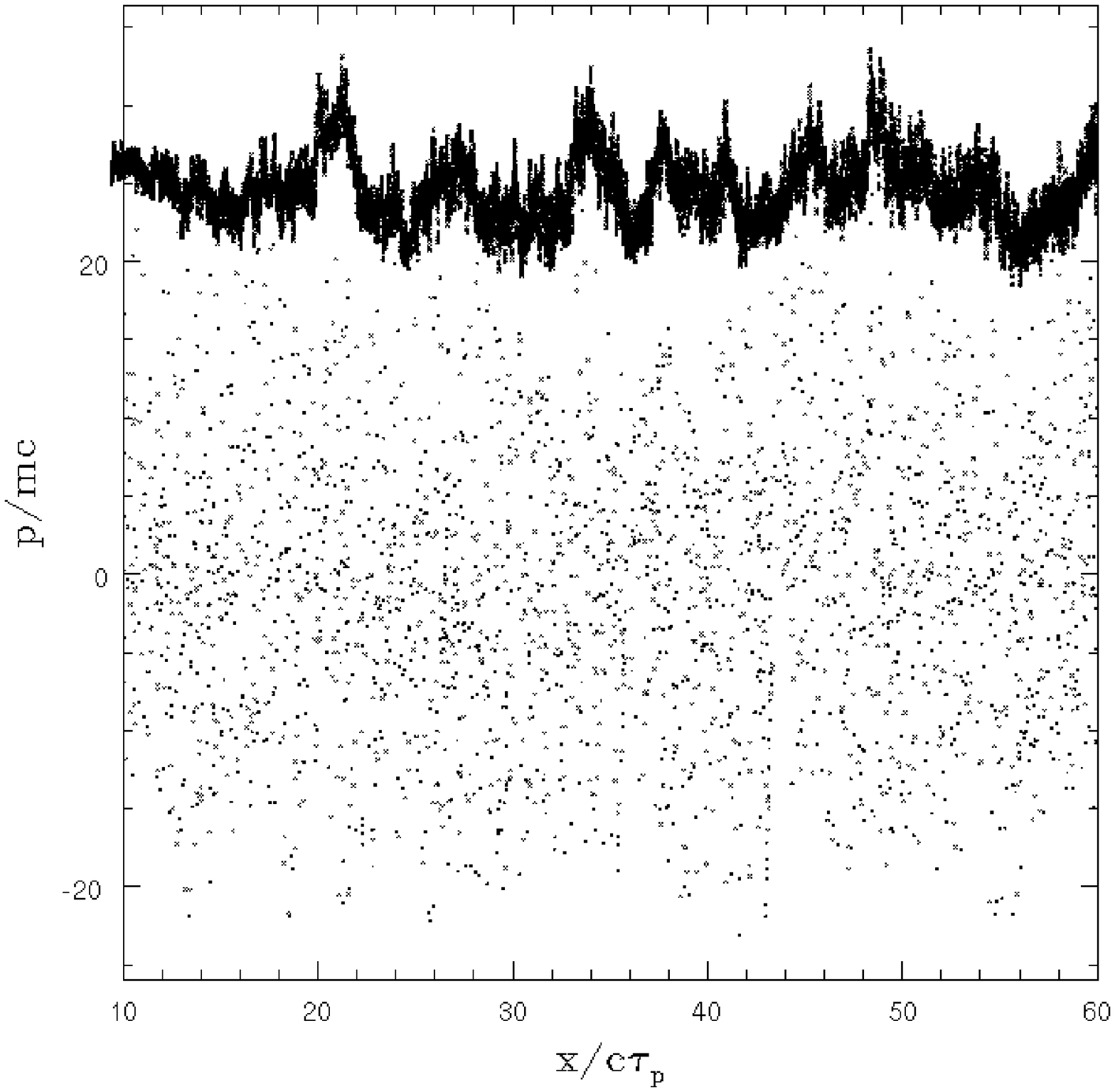,width=6cm,clip=} 
\psfig{file=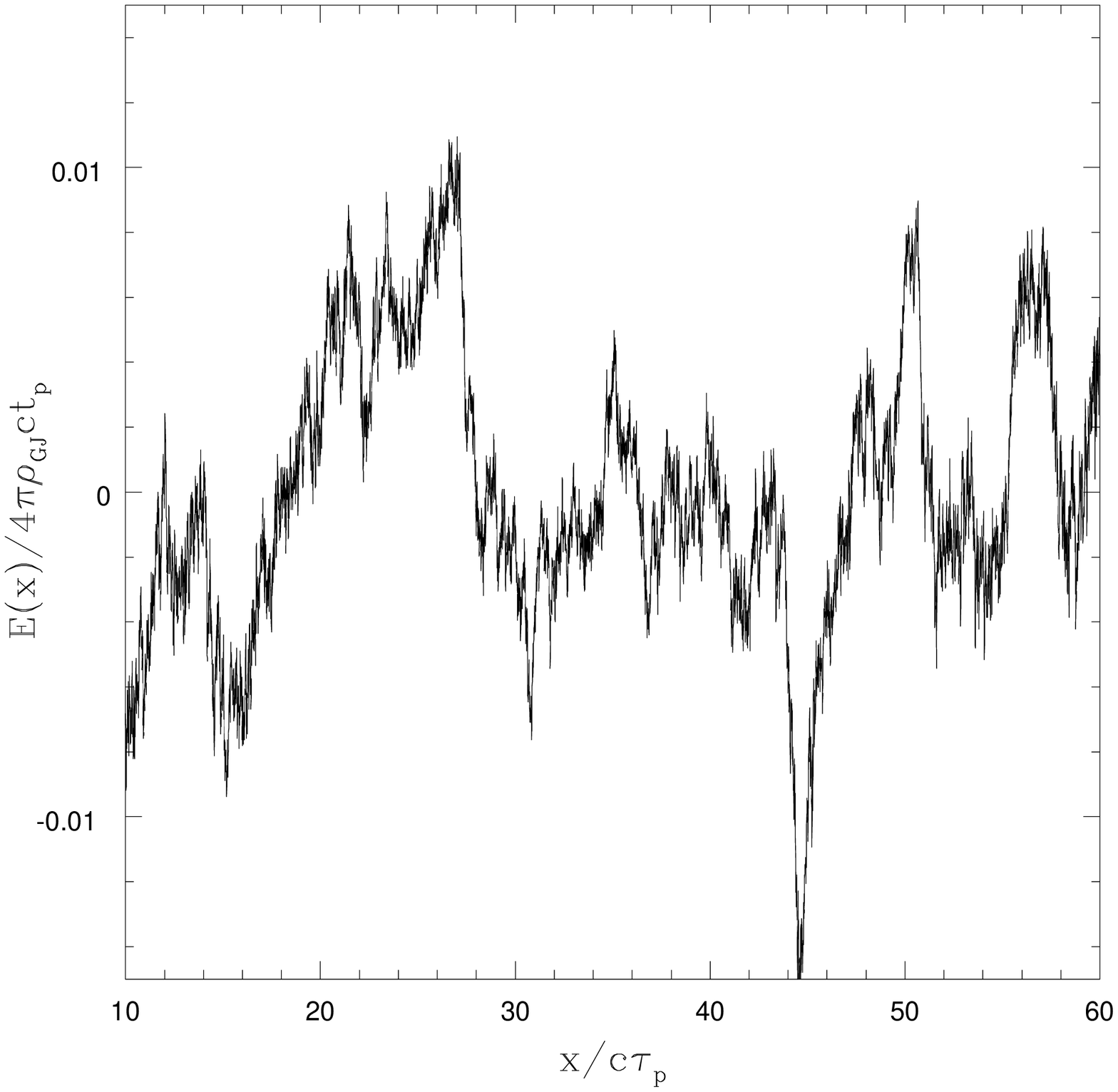,width=6cm,clip=} }
\caption{Excitation of strong waves in a sub-GJ current. In this example,
which is typical of our numerical experiments, a current
density $j = 0.95 j_{\rm GJ}$ flows from the surface of the star ($x=0$
on the grid).  The simulation follows the positions and momenta of 
individual charges, and the electric field evolution in time and space. 
Left, a partial time series of the $E$ field at one point of the
grid.
Center, a phase plot showing the resultant particle
trapping.  Right, a shapshot of the strong 
waves in the electric field, at one time during the run.  We find
that the frequency of the waves excited decreases as $j / j_{\rm GJ}$ 
approaches unity. 
\label{image}}
\end{figure*}

Numerical simulations 
are needed to follow the plasma evolution and determine the time
signature of each model.  Weatherall (1997, 1998) has carried
out such simulations for the plasma turbulent emission model. 
In this picture, a two-stream
instability drives electrostatic plasma turbulence.   When the
turbulence becomes nonlinear, the modulation instability produces
filamentation of the initial beam-resonant wave, which is followed
by ``collapse'' of the wave along the field direction.  Thus,
the wave energy  accumulates in transient, spatially localized 
regions. Figure 5 shows an example of this
growth and collapse.  Mode conversion during the explosive collapse
phases generates radio emission (the coherence comes from the spatial 
localization of the plasma oscillation).
The simulations reveal  a distinctive time signature (Figure 6 
shows one example)  which comes from the coupling of
the radiation modes to the beam-resonant wave.  The 
simulations predict quite narrow band emission, $\Delta \nu / \nu \sim 0.2$,
centered on the {\it comoving} plasma frequency;  and predict
timescales given by $\nu \delta t \sim O(10)$ (thus, a few ns at 5 GHz).

Other competing emission mechanisms will
have their own characteristic timescales and temporal signatures
associated with the ``shots''. 
For instance, curvature emission from coherent plasma bunches
might evolve on a particle trapping timescale, \eg, $( m \lambda
/ e E)^{1/2} $, where $\lambda$ is the bunch length and the
field is limited by $E^2 \ltw 8 \pi \gamma_{\rm p} m c^2 n_{\rm p}$.  In addition,
the bunch life may be limited by magnetic field curvature
(\eg, Melrose 1992).  Combining these, we estimate this process has
a longer timescale, $\sim 0.01 - 0.1 \mu$s;  but numerical simulations
are required to confirm such speculation.

As the emission region is very likely inhomogeneous, 
containing many microscopic emitting structures (growing and 
collapsing wave packets, or coherent bunches which grow and
dissipate), an
observed pulse will be an incoherent collection of such nanobursts.
Each nanoburst will contibute an impulse in a shot noise-type model.  
The
individual impulses may be observable if their density is low enough;
however a full understanding of the pulse nanostructure requires
an understanding of the statistics of such shots. This 
also requires numerical simulation, and we are extending our
plasma-turbulence work to address this issue.

\subsection{Meso-scale: dynamical fluctuations.}

When we consider the intermediate timescales of 
microstructure, we have likely  left the regime of the
fundamental radiation process. 
The associated length scales are comparable to the dimensions of
the region (the acceleration region or the radio emission region). 
Thus, we are ``seeing'' the dynamical behavior of the radio-loud plasma,
as it responds to local forces. 

As an example of mesoscale dynamics, consider plasma flow in the
acceleration region.  The
general $\bold E$ field in this region, whose divergeance is the
physical charge density $\rho$, can be separated into
``corotation'' and ``other'' parts: 
\be
{\tilde \mathbf E} =  {\mathbf E} - {\mathbf E}_{\rm co} ~; 
\quad
\del \cdot 
{\tilde \mathbf E} = 4 \pi \left( \rho - \rho_{\rm GJ} \right)
\ee 
The easiest way to follow plasma dynamics in this region is to
work in the rotating frame, where the difference field,
$\tilde {\mathbf E}$, controls particle motion.   (Compare equation 1,
where ${\mathbf E} = {\mathbf E}_{\rm co}$ was assumed).  Thus, if ${\mathbf E}
\ne {\mathbf E}_{\rm co}$ (that is if $\rho \ne \rho_{\rm GJ}$), the plasma
feels a net force in the corotating frame.

Now, let the  acceleration region sit within a pulsar circuit, 
so that the local
current density is determined by the voltages and resistive loads of
the full circuit.  This density can easily differ from GJ.  If it
exceeds GJ (in absolute value), simple space-charge-accelerated flow
can occur.  If it is somewhat below GJ, however, the situation becomes
more complicated.  Several authors (\eg, Shibata 1997, or Jessner \etal,
 this proceedings) have
noted that steady-flow solutions display strong spatial oscillations when
the current is sub-GJ.  

We revisit this problem, allowing for time variability. We use a
one-dimensional particle-in-cell code to model the space-charge
acceleration of plasma escaping from the star's surface.  
The simplest
situation starts with an external vacuum and a background (corotation)
field.
For super-GJ flow we verify that
the system reaches the steady, space-charge-limited situation,
$\Phi(z)  \propto \gamma(z) \propto z^2$ (where $\Phi$ is the potential
and $z$ is the space coordinate; described \eg\/ by
Shibata). In the interesting case of slightly sub-GJ flow, we also
find oscillations, but now they are temporal as well as
spatial.  The outflowing plasma
rapidly develops strong waves in the density and $E$ field, in which
feedback and nonlinear growth tends to trap the particles between
the wave crests.  This trapping slows, but does not stop, the advance of
the charge front.  A quasi-steady state is reached, in which the
current is modulated by strong ``trapping'' waves.  
Figure 7 shows an example of this flow.  Shibata predicted the spatial
wave frequency would depend on the current density;  we find this
is true also of the temporal frequency.  As the current 
approaches the corotation value, the oscillations slow down.  
We have found oscillation periods $\sim 10 - 100 \nu_{\rm p}^{-1}$ 
in our simulations
to date, which  could be observable as microstructure
if $\nu_{\rm p}$ is identified with the observing frequency.

Our simulations do not include emission of escaping radiation.  Thus,
we cannot directly predict the time signature of the coherent
radiation emitted by this plasma.  That radiation will arise through
the interaction of plasma charges and waves riding within the meso-scale
structure which these simulations describe.  We anticipate, however, 
that the density and field modulations seen in this non-stable system will
be echoed in the escaping radiation.  Thus, pulsars with strong
microstructure signatures may be ones in which the current is close
to but below the corotation current.

\begin{figure*}[htb]
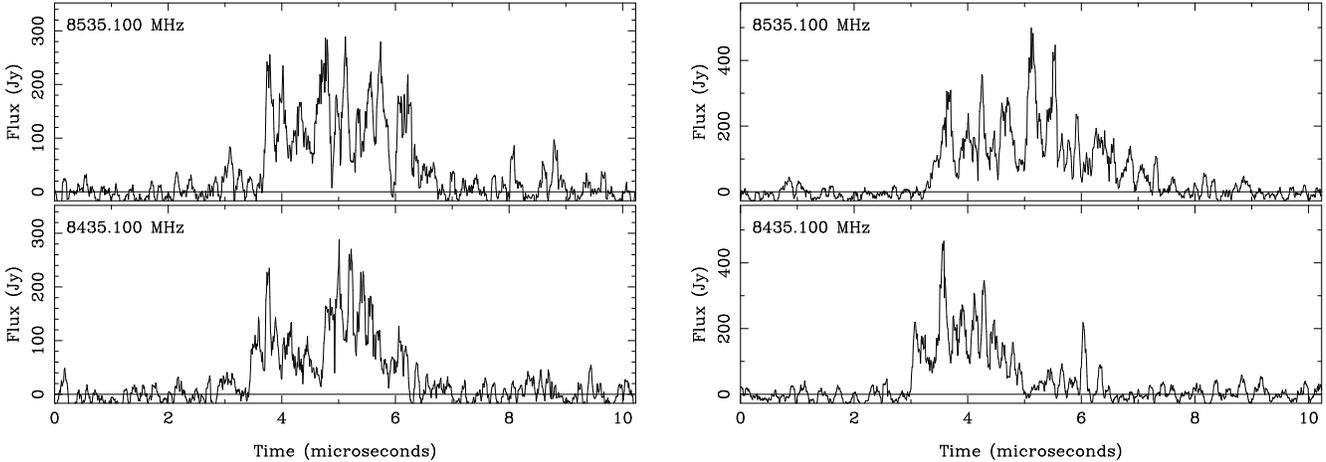

\centerline{
{\rotatebox{-90} 
{\resizebox{2.4in}{!}
     {\includegraphics{eilek_fig8a.ps}}}}
\hspace{0.2in}
{\rotatebox{-90} 
{\resizebox{2.4in}{!}
     {\includegraphics{eilek_fig8b.ps}}}}
}
\caption{Two single giant pulses from the interpulse of the
Crab pulsar, each recorded
simultaneously at 8.4 and 8.5 GHz.  The data were obtained (after
coherent dedispersion; \S10.1) at 10 ns time resolution, and
have been block smoothed to 80 ns.
 The dispersion delay has been removed. 
\label{image}}
\end{figure*}

We expect similar results in other situations. 
One example is pair formation, which might
disrupt an initially steady flow.  Laboratory diodes 
are difficult to
maintain in a steady state, because plasma formation at the diode
disrupts the current, leading to strong waves and breakdown of the
steady flow.  We expect that pair formation in a pulsar acceleration 
region will have a similar effect.  Rapid injection
of new plasma in a steady-flow region will disrupt the flow, leading
to temporal variability that may be observable. 

\subsection{Macro-scale:  self organization.} 

On longer timescales, many rotation periods, we leave plasma
microphysics and go to larger-scale, MHD effects.  These timescales
are the most accessible observationally, and perhaps the most
daunting theoretically.

One striking example is the long-lived drifting subpulses seen in
some stars.  Their drift rate is consistent with the ${\mathbf E}
\times {\mathbf B}$ drift due to the  non-corotation $\bold E$ 
field predicted by well-known models
of the acceleration region (\eg, Ruderman \& Sutherland, 1975,
 or Arons \& Scharlemann, 1978). 
The mystery of the drifting subpulses is their stability;  the
analysis of Deshpande \& Rankin (2000) suggests that individual
radio-loud structures retain their identity for 10--100 rotation
periods.  They have been modelled as ``sparks''
of localized pair creation, which may partially shield their
immediate surroundings (\eg, Ruderman \& Sutherland).
  This does not seem to be the
entire picture, however.  Why should such sparks be so long-lived?

It is tempting to speculate that these long-lived plasma
filaments are the result of a saturated plasma instability.  Such
features -- for instance, quasi-stable,
current-aligned filaments -- are common in other plasma environments.  
The problem is how to make them.  The radio-loud pulsar plasma is
very highly magnetized, as shown by the energy ratio 
\be
\beta = {8 \pi \gamma n m c^2 \over B^2 } \sim 1 \times 10^{-18} { \gamma n 
\over n_{\rm GJ}} { \left( { r/ r_*} \right)^3 \over P B_{*,12}} 
\ee
(we have scaled the number density to the corotation value, the
$B$ field to its value at the star's surface, and assumed a dipole field).
This shows that one must go to quite high altitudes before the plasma inertia
becomes comparable to the magnetic energy density (so that normal MHD
instabilities might be expected).  Resistive instabilities are tempting;
they are known to occur in low-$\beta$ plasmas and can lead to field-aligned
filaments.  We agree with  Melrose (2000) who
 suggested such structures in the context
of the density problem.  It may be that filaments 
exist in all pulsars at some level,
and happen to be stronger or longer-lived in those with clear subpulse
drift.  However, without a good understanding of such details as
cross-field resistivity in this environment, we can do no more than
speculate on this interesting question.

\begin{figure*}[htb]
\centerline{
{\rotatebox{-90} 
{\resizebox{2.4in}{!}
     {\includegraphics{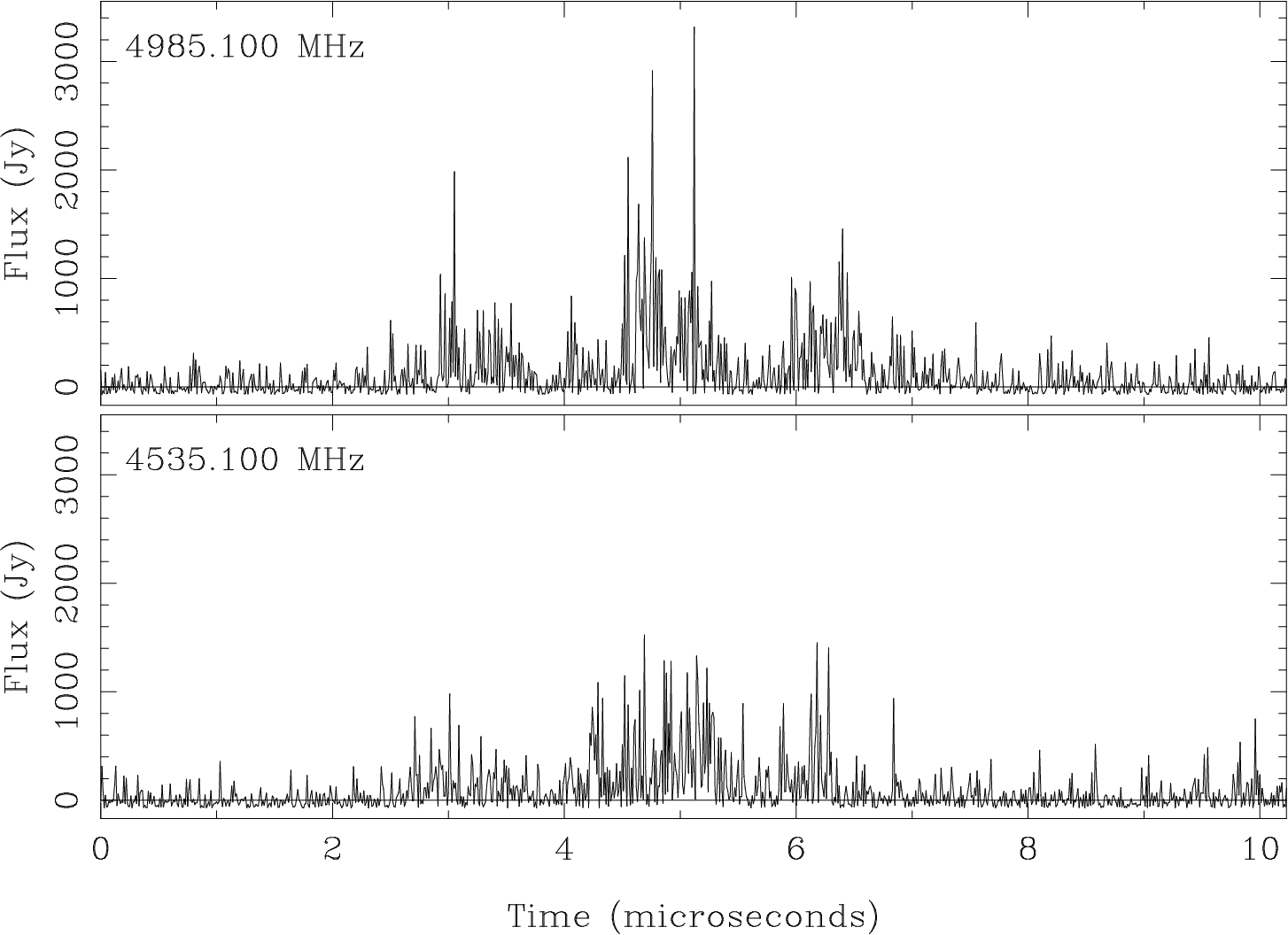}}}}
\hspace{0.5in}
{\rotatebox{-90} 
{\resizebox{2.4in}{!}
     {\includegraphics{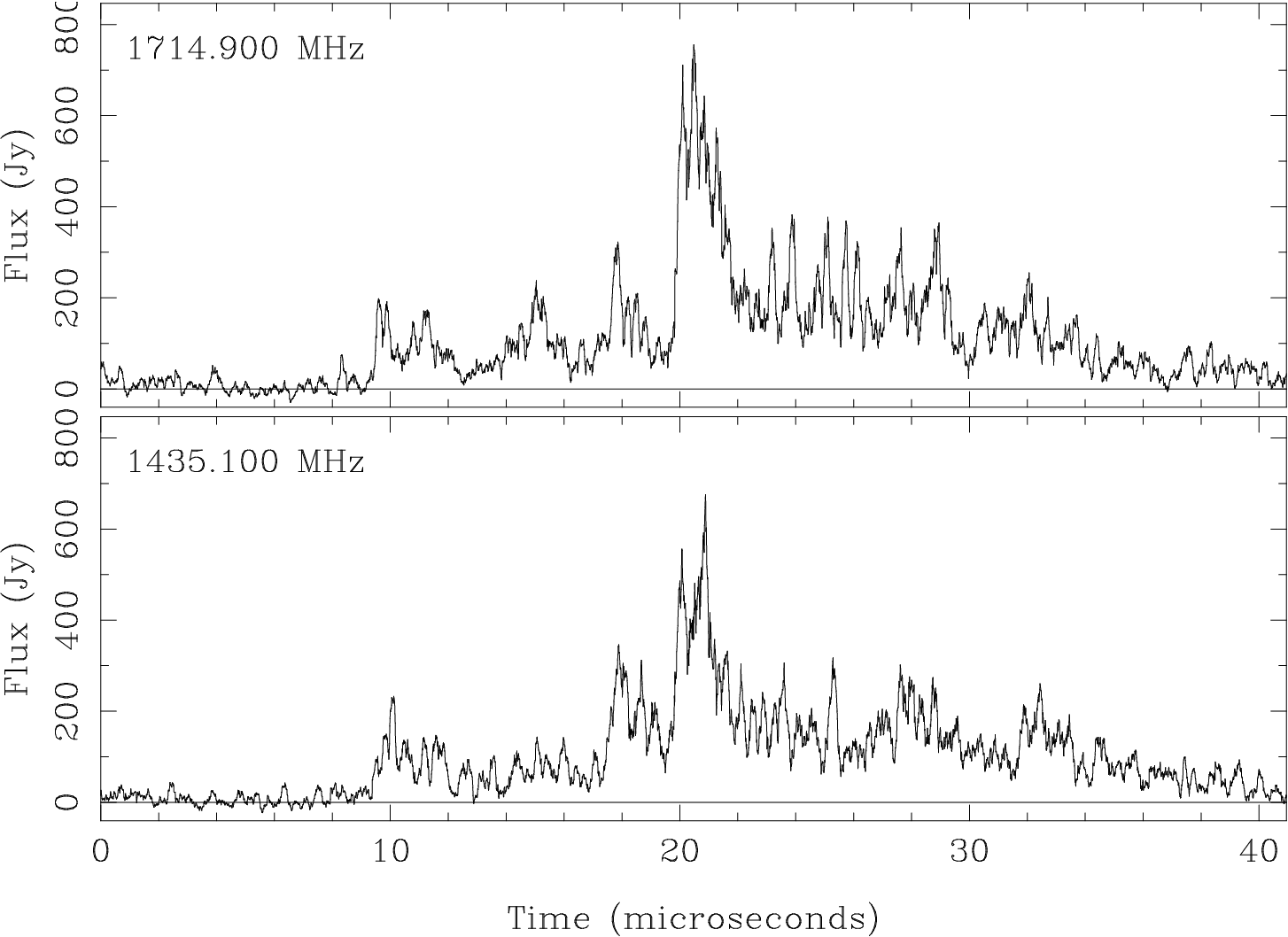}}}}
}
\caption{Two single giant pulses from the Crab pulsar.  Left, a pulse
recorded simultaneously at two frequencies within C band, with time resolution
10 ns.  Right, a (different) pulse recorded simultaneously at two
frequencies within L band;  the data have were obtained at 10 ns time
resolution and have been smoothed to 200 ns.  The dispersion delay has
been removed for each pulse.  These pulses illustrate
the typical behavior of the star.  The subpulses and micro-bursts are
similar but not identical. Such data suggest that the 
intrinsic bandwidth of an individual ``burst'' of emission is at
least  $\delta \nu  / \nu  \sim  0.2$.
\label{image}}

\centerline{
{\rotatebox{-90} 
{\resizebox{2.4in}{!}
     {\includegraphics{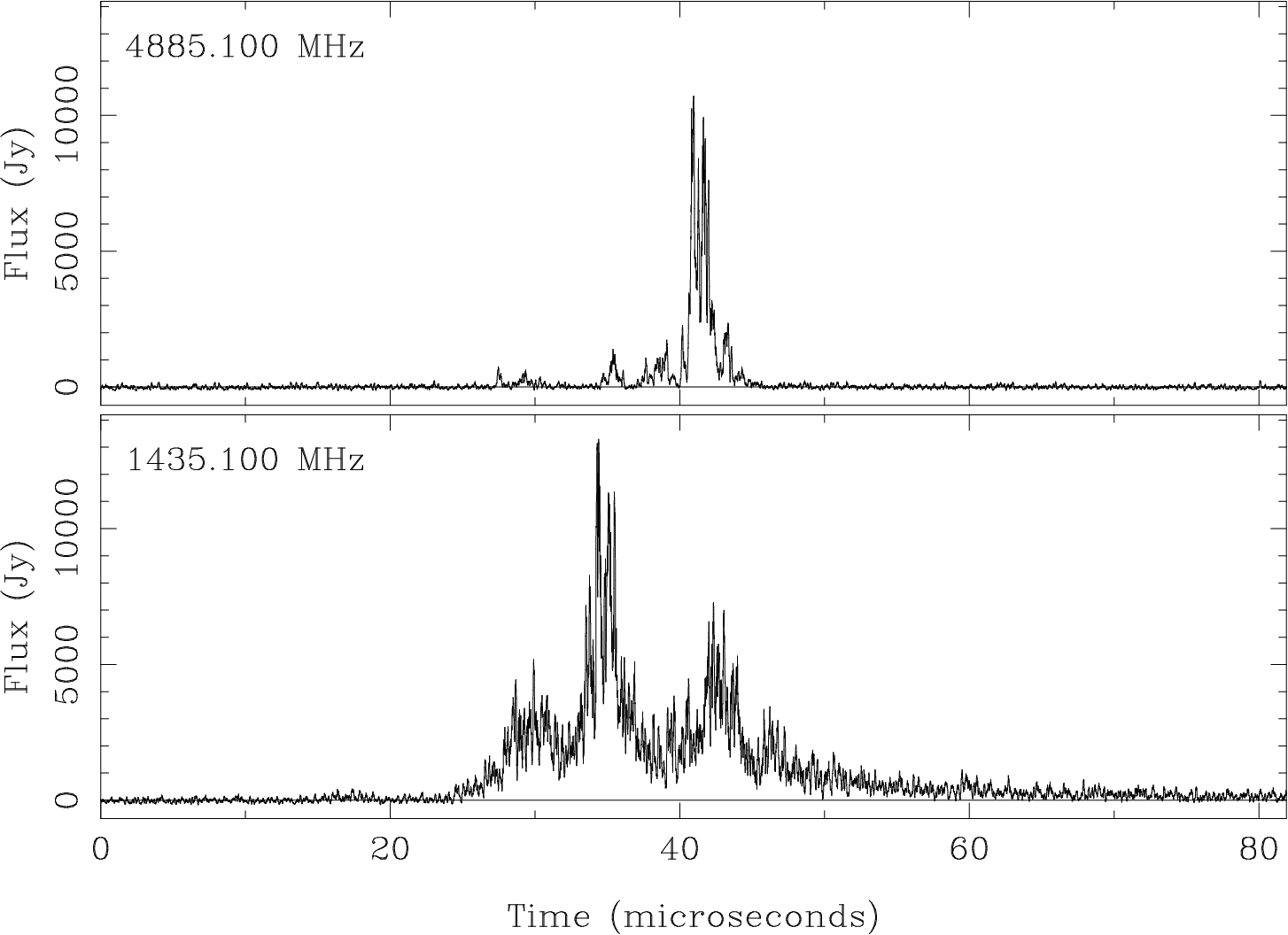}}}}
\hspace{0.2in}
{\rotatebox{-90} 
{\resizebox{2.4in}{!}
     {\includegraphics{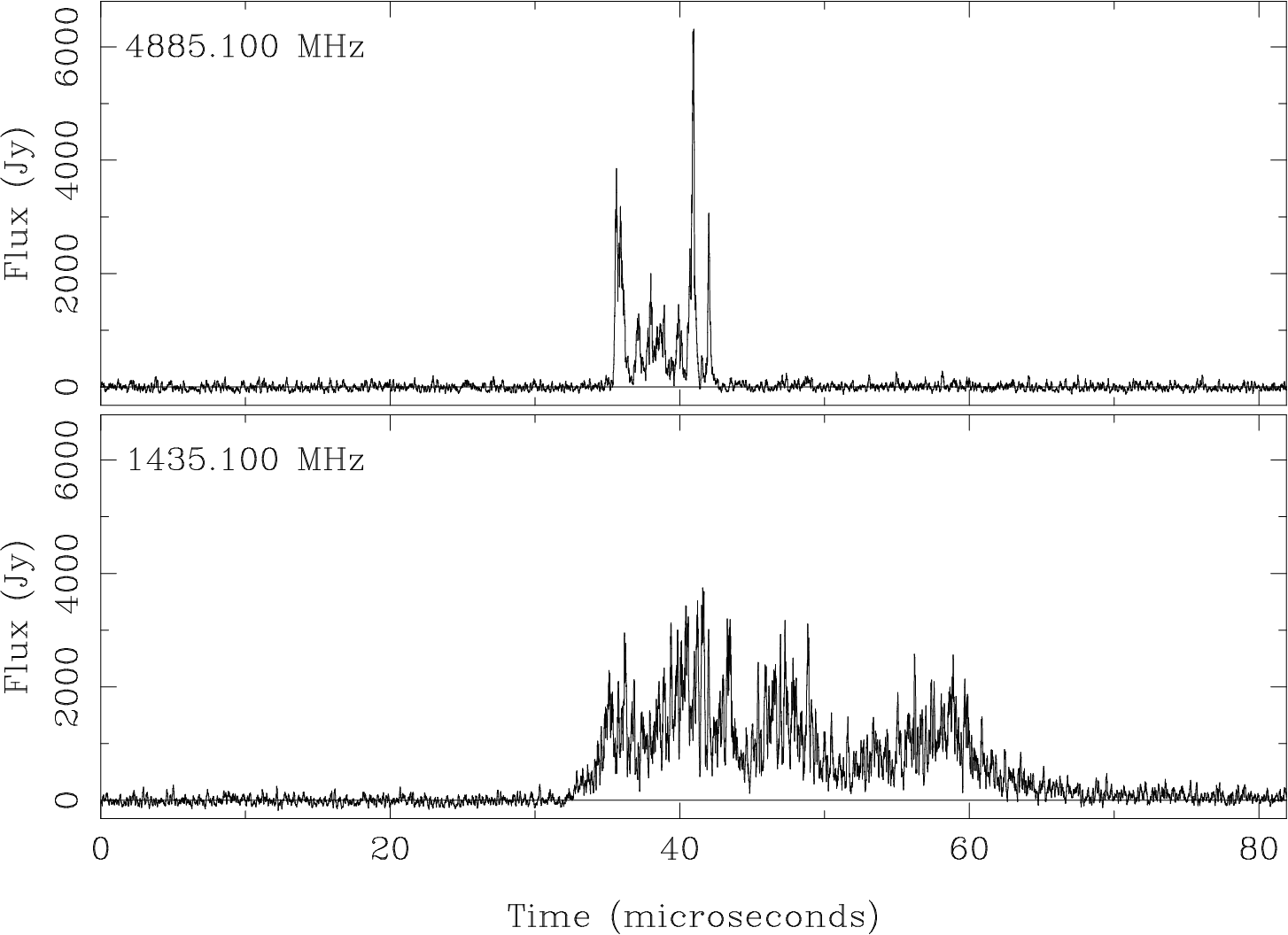}}}}
}
\caption{Two more single giant pulses from the Crab pulsar.  Each
was recorded simultaneously at 5 and 1.4 GHz, obtained at 10 ns time
resolution and smoothed to 100 ns.  The dispersion delay has been
removed.  These two pulses are not as well correlated, over
this larger frequency separation, as are the two illustrated in Figure 9.
This behavior is typical of the star.  We infer that the emission bandwidth
of the microbursts which constitute an individual ``giant pulse'' is
narrower than this frequency separation.  Comparing this with data
such as in Figure 9, we infer the intrinsic emission bandwidth obeys
$0.2 \ltw \Delta \nu / \nu < 1$,  and that a giant pulse contains a
collection of many such microbursts, distributed over a range of
central emission frequencies. 
\label{image}}
\end{figure*}

\section{A Case Study:  Giant Pulses in the Crab pulsar}

The Crab pulsar emits occasional ``giant''
pulses which can be thousands of times more intense than an average 
pulse. One of us (Hankins, with his colleagues) is making
 ultra-high time resolution observations of these giant pulses.
The data can give us insight into the emission mechanism and the 
dynamics of the emitting plasma. In this section we
 present preliminary results which bear on the physics of the
 emission region. 

\subsection{Observations and Pulse Structure} 

Using the VLA in phased-array mode minimizes the signal from 
the surrounding nebula, and allows us
to study individual giant pulses at time resolution down to a
few ns.  To achieve this we used coherent dedispersion.  The
incoming voltage is Nyquist sampled at 100 MHz.  The samples
are recorded for later, off-line processing which removes
dispersion distortion.  The signals are passed through a filter
whose transfer function is the inverse of the (frequency-dependent)
phase change imposed by propagation through the ISM.  This technique
allows us to recover the full time resolution of the emitted signal,
limited only by the receiver bandwidth. 

We reached 10 ns resolution
 at several frequencies from 1.4  to 15 GHz, making 
single frequency observations and also doing
simultaneous observations at two frequencies.  
Individual pulses are quite complex when seen at this resolution.
They often contain clear subpulses, which usually have
 a ``Fred'' shape (``{\em F}ast {\em R}ise, 
{\em E}xponential {\em D}ecay'').   Figures 8 - 11 show some examples.
The subpulse widths are typically $\sim 1 \mu$s at 5 GHz, and
$\sim 0.1 \mu$s at 8 GHz.  (We note that  unresolved features can be 
found down to our resolution limit of 10 ns.)

\begin{figure*}[htb]
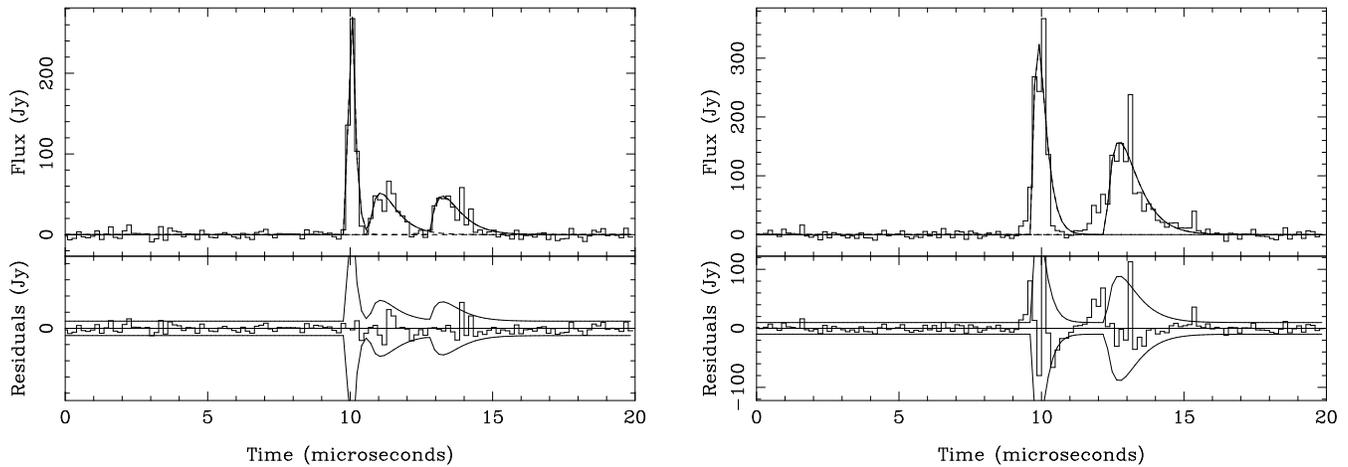

\centerline{
\rotatebox{-90}{\resizebox{2.4in}{!}
     {\includegraphics{eilek_fig11a.ps}}}
\hspace{0.2in}
\rotatebox{-90}{\resizebox{2.4in}{!}
     {\includegraphics{eilek_fig11b.ps}}}
}
\caption{Two representative X band pulses, illustrating ``Fred'' fits
to subpulses.  The data were 
fitted with  a function $I(t) \propto I_o t e^{-t/\tau}$. (Thus, $I_o$
measures the pulse height, $\tau$ the width, and the total energy 
$\propto I_o \tau$.)
The upper plot shows the original data, taken at 8415 MHz and
smoothed to 160 ns resolution, and the result of the ``Fred'' fits. 
 The lower plot shows the $2 \sigma$ estimation error of the fit and the
fit residuals.  The intensity
$I_o$ and width $\tau$ for a set of these fits are plotted in Figure 12.
}
\end{figure*}

Above $\sim 2$ GHz we clearly see the intrinsic pulse shape;
below $\sim 1$ GHz the pulse width is dominated by nebular or
interstellar 
scattering (ISS). In the range $\sim 1 - 2$ GHz 
single pulses or subpulses show the ``Fred'' shape characteristic
of scattering, but have widths which do not follow the $\nu^{-4}$ 
law expected for ISS, and  are  strongly variable
on a timescales less than an hour (too short to be from variations in
the ISM).
 The average pulse width  at high frequencies
decays only slowly with frequency, and may obey $\Delta t (\nu) \propto
\nu^{-2}$ (Moffett, 1997).
 We suspect that propagation through the magnetosphere 
broadens high-frequency subpulses and give them the ``Fred'' shape.  
Turbulent scattering in a cold, relativistically
streaming pair plasma in fact gives the observed frequency signature
(Eilek, 2002).

\subsection{Emission bandwidth} 

Giant pulses in the Crab pulsar occur simultaneously over a
broad frequency range, at least from 1.4 to 4.8 GHz.  Thus, in
some sense the emission is broad-band.

However, the situation becomes more complex if we consider 
substructure within a giant pulse.  Our simultaneous two-frequency
data suggest that the intrinsic emission bandwidth of {\it substructure} is
narrow band.  Compare Figures 8-10 in which we present
simultaneous dual-frequency measurements of individual giant pulses.
While we have not yet completed a statistical analysis of our two-frequency
observations, our impression after working with the data is that
the substructure in a giant pulse is well correlated if
the two frequencies are close:  1.4 and 1.7 GHz, 4.5 and 5.0 GHz,
or 8.4 and 8.5 GHz.  However, the substructure is less well
 correlated over a broader
frequency range, such as 1.4 and 4.8 GHz. This suggests that the
bandwidth of a given {\it subpulse} obeys $0.2 \ltw \Delta \nu / \nu < 1$.
Comparing this to the simulations by Weatherall (1998), we note that
the observations are consistent with what his model predicts.

\subsection{Energy content}

Another interesting hint in the data is
 that individual subpulses conserve total energy, $E = \int I(t) dt$.  
We have fitted several data sets at 1.4, 4.8 and 8.4 GHz with the 
``Fred''-like function 
$I(t) = I_o t e^{-t/\tau}$, allowing multiple subpulses per giant
pulse. Figure 11 shows examples of such fits.
  Typical results are shown in Figure 12.  A plot of intensity $I_o$
against width $\tau$ 
shows an upper envelope, which is consistent with $E \propto
I_o \tau =$ constant.  This is what one would expect if each subpulse
results from a ``collect and release'' process, in which internal
energy is stored until a threshold is
reached, then suddenly released.  The constant-energy envelope of
the distribution then  describes the total energy released in the
burst, and the points below the envelope are what we would expect if
the released energy  is relativistically
beamed at some angle to the line of sight.  

\begin{figure}[htb]
\centerline{\psfig{file=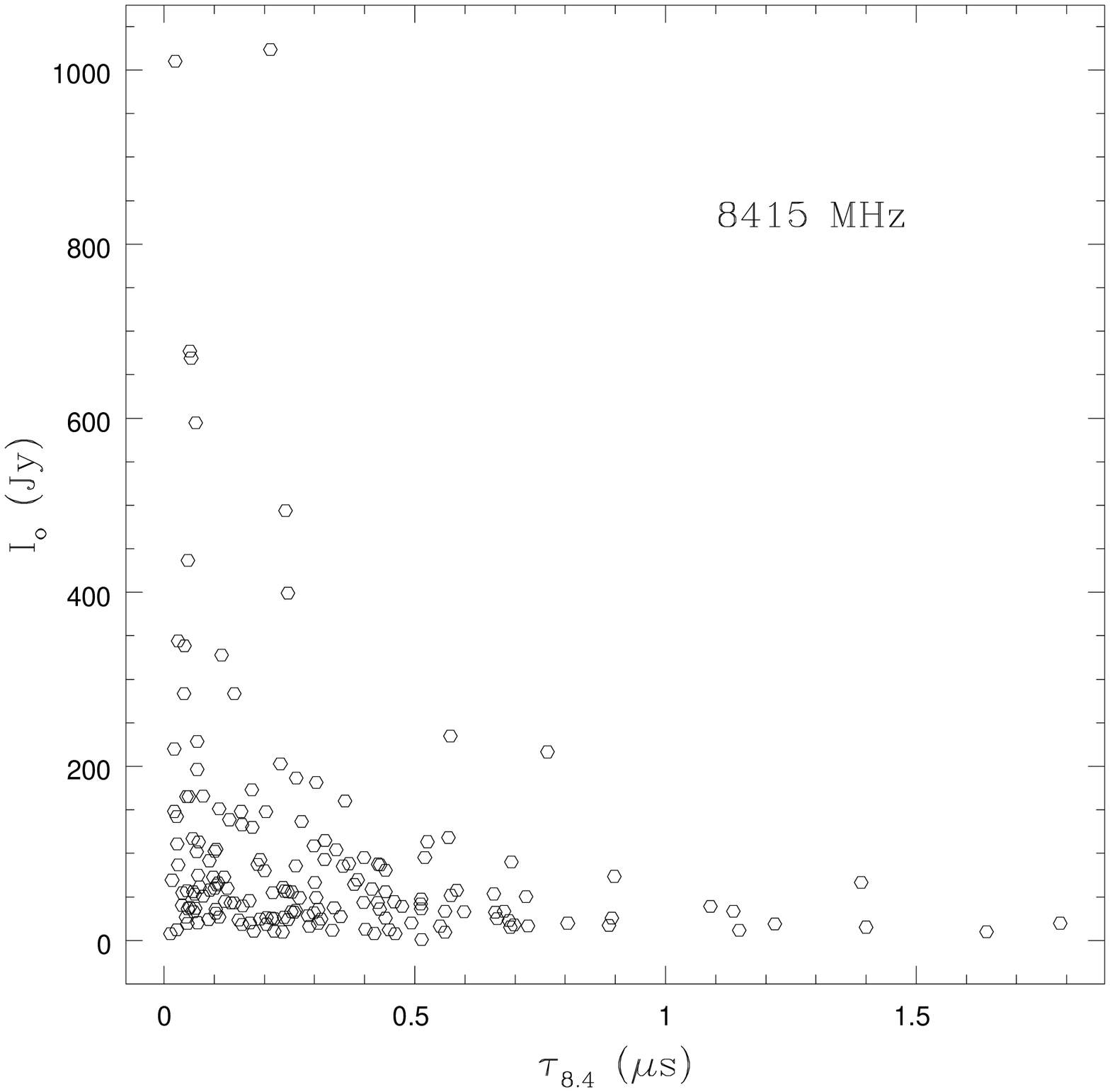,width=6.8cm,clip=} }
\caption{Intensity {\it vs.}  width for subpulse fits for one data
set.  The plot shows $(\tau, I_o)$ from fits to 
180 subpulses identified in a set of 60 giant pulses, 
recorded within a bit over an hour at 8.4 GHz
at the VLA.  There is a tendency for
weak pulses to be broad and strong pulses to be narrow;  a constant
energy ($I_o \propto 1 / \tau$) envelope seems to exist.
\label{image}}
\end{figure}

We caution the reader that these results are preliminary, and require
a careful statistical analysis of a larger, complete sample of
high-frequency subpulses (currently underway).   If further work
supports our speculation, that the subpulses in giant pulses have
an upper limit to their energy content, then the difference between
this conclusion and the energy distribution of giant pulses
presented by Lundgren 
\etal\/ (1995) must be addressed.  We note two differences between our
work and theirs.  We are considering
subpulses, several of which may be contained in a giant pulse.  In
addition, 
our higher time resolution allows us to separate subpulse width and
height, which may make the upper energy limit more apparent.

\subsection{The emission region in the Crab pulsar}

From these data, we begin to draw a picture of the radio-loud
plasma in the Crab pulsar.  It is inhomogeneous and fluctuates on
quite short timescales.    Any given piece of the plasma is a
narrow-band radio emitter;  the larger frequency range of an
individual giant pulse probably comes from the variable density 
within the region.   A giant pulse seems to trace
the simultaneous excitation of an extended region of plasma, in
which different sub-regions independently reach some critical 
threshold and then release their stored energy in sudden, observable
bursts. 

\vspace{0.12in}

{\bf Acknowledgements}.  We have learned a great deal 
about pulsar physics through discussion with a number of people, 
including Jon Arons, Gregory Benford,
Janusz Gil, Wolfgang Kundt, Harald Lesch, Leon Mestel, Don Melrose
and Dipanjan Mitra. We particularly appreciate the enthusiastism
and hard questions supplied by Jeff Kern and Axel Jessner, and we
thank Axel for his careful reading of this long paper.
Work by the Socorro pulsar group
has been partially supported by NSF grants AST-9315285 and AST-9618408.
JE appreciates support and hospitality from the Heraeus foundation
and the MPIfR while this paper was being written.


\newpage

\begin{thebibliography}{} 

\bibitem{}Arendt, P. N. Jr, 2002, Ph.D. thesis, New Mexico Tech

\bibitem{}Arendt, P. N. Jr. \& Eilek, J. A., 2002, submitted to ApJ.

\bibitem{} Arons, J., 1981, in Proc. Varenna Summer School on Plasma
Astrophysics, ed. T. D. Guyenne \& G. L\'evy (Noordwijk:  ESA), 273

\bibitem{} Arons, J. \& Scharlemann, E. T., 1979, ApJ, 231, 854

\bibitem{} Asseo, E., Pelletier, G. \& Sol, H., 1990, MNRAS, 247, 529

\bibitem{} Eilek, J. A., 2002, submitted to ApJ

\bibitem{} Eilek, J. A. \& Hankins, T. H., 2000, in IAU Colloquium 177, 
Pulsar Astronomy--2000 and Beyond,  eds., M. Kramer, N. Wex 
\& R. Wielebinski (San Francisco: ASP), 721 

\bibitem{} Eilek, J. A. \& Hankins, T. H., 2002, in preparation

\bibitem{} Daughtery, J. K. \& Harding, A. K., 1982, ApJ, 252, 337

\bibitem{} Deshpande, A. A. \& Rankin, J. M., 2001, MNRAS, 322, 438


\bibitem{} Hankins, T. H. \& Rankin, J. R., 2002, in preparation. 

\bibitem{} Hibschman, J. A. \& Arons, J., 2001, ApJ, 560, 871

\bibitem{} Jessner, A., Lesch, H. \& Kunzl, T., 2001, ApJ, 547, 959

\bibitem{} Kazbegi, A. Z., Machabeli, G. Z. \& Melikidze, G. I.,
1995, MNRAS, 253, 377

\bibitem{} Kijak, J. \& Gil, J., 1997, MNRAS, 288, 631

\bibitem{}Kunzl, T., Lesch. H., Jessner, A. \& von Hoensbroech, A.,
1998, ApJ, 505, L139

\bibitem{}Kunzl, T., Lesch, H. \& Jessner, A., 2002, submitted to
A\&A

\bibitem{} Lange, C., Kramer, M., Wielebinski, R. \& Jessner, A., 1998,
A\&A, 332, 111

\bibitem{} Lundgren, S. C., Cordes, J. M., Ulmer, M., Matz, S. M.,
Lomatch, S., Foster, R. S. \& Hankins, T. H., 1995, ApJ, 453, 433

\bibitem{} Luo, Q. \& Melrose, D. B., 1995, MNRAS 276, 372

\bibitem{} Lyutikov, M. \& Parikh, A., 2000, ApJ, 541, 1016.

\bibitem{} Melrose, D. B., 1992, in IAU Colloquium 128, ed. T. H.
Hankins, J. M. Rankin \& J. A. Gil (Zielona G'ora:  Pedaogical Univ.
Press), 307

\bibitem{}Melrose, D. B., 2000, in IAU Colloquium 177, 
 eds., M. Kramer, N. Wex  \& R. Wielebinski (San Francisco: ASP), 721

\bibitem{} Moffett, D., 1997, Ph.D. thesis, New Mexico Tech

\bibitem{} Petrova, S. A. \& Lyubarskii, Y. E., 2000, A\&A, 355, 1168

\bibitem{} Ramachandran, R., Rankin, J. M., Stappers, B.W., Kouwenhoven,
M.L. A., \& van Leeuwen, A. G. J., 2002, submitted to A\&A

\bibitem{} Rankin, J. R., 1993, ApJ, 405, 285

\bibitem{} Rathnasree, N. \& Rankin, J. R., 1995, ApJ, 452, 814

\bibitem{} Ruderman, M. A. \& Sutherland, P. G., 1975, ApJ, 196, 51

\bibitem{} Shibata, S., 1991, ApJ, 278, 239

\bibitem{} Shibata, S., 1997, MNRAS 287, 262

\bibitem{} Shibata, S., Miyazaki, J. \& Takahara, F., 1998, MNRAS 295, L53

\bibitem{} Sturmer, S., 1995, ApJ, 446, 292

\bibitem{} van Hoensbroech, A. \& Xilouris, K. M., 997, A\&A, 126, 121

\bibitem{} Weatherall, J. C., 1994, ApJ, 428, 261

\bibitem{} Weatherall, J. C., 1997, ApJ, 483, 402

\bibitem{} Weatherall, J. C., 1998, ApJ, 506, 341

\bibitem{} Weatherall, J. C., 2001, ApJ, 559, 196

\bibitem{} Weatherall, J. C. \& Eilek, J. A., 1997, ApJ, 474, 407



\end{thebibliography}
\end{document}